**Title: Electric-field-driven spin resonance by on-surface exchange coupling to a single-atom magnet**
**Short title: Spin resonance driven by coupling to a single-atom magnet**

**Authors:**
Soo-hyon Phark[1,2,3]†, Hong T. Bui[1,4]†, Alejandro Ferrón[5], Joaquin Fernández-Rossier[6], Jose Reina-Gálvez[1,2], Christoph Wolf[1,2], Yu Wang[1,2], Kai Yang[3,7], Andreas J. Heinrich[1,4]\*, Christopher P. Lutz[3]\*

**Affiliations:**
[1]Center for Quantum Nanoscience, Institute for Basic Science (IBS), Seoul 03760, Korea
[2]Ewha Womans University, Seoul 03760, Korea
[3]IBM Research Division, Almaden Research Center, San Jose, CA 95120, USA
[4]Department of Physics, Ewha Womans University, Seoul 03760, Korea
[5]Instituto de Modelado e Innovación Tecnológica (CONICET-UNNE) and Facultad de Ciencias Exactas, Naturales y Agrimensura, Universidad Nacional del Nordeste, Avenida Libertad 5400, W3404AAS Corrientes, Argentina
[6]International Iberian Nanotechnology Laboratory (INL), 4715-330 Braga, Portugal
[7]Beijing National Laboratory for Condensed Matter Physics and Institute of Physics, Chinese Academy of Sciences, Beijing, China.

\*Corresponding authors. Email: heinrich.andreas@qns.science, cplutz@us.ibm.com
†These authors contributed equally to this work.

**Abstract:**
Coherent control of individual atomic and molecular spins on surfaces has recently been demonstrated by using electron spin resonance (ESR) in a scanning tunneling microscope (STM). Here we present a combined experimental and modeling study of the ESR of a single hydrogenated Ti atom that is exchange-coupled to a Fe adatom located in 0.6-0.8 nm away. Continuous wave and pulsed ESR of the Ti spin showed a Rabi rate with two contributions, one from the tip and the other from the Fe, whose spin interactions with Ti were modulated by the radio-frequency electric field. The Fe contribution is comparable to the tip, as revealed from its dominance when the tip was retracted, and tunable using a vector magnetic field. Our new ESR scheme allows on-surface individual spins to be addressed and coherently controlled without the need for magnetic interaction with a tip. This study establishes a feasible implementation of spin-based multi-qubit systems on surfaces.

**Teaser:**
Electron spin resonance of a single atom is driven by exchange coupling to a single-atom magnet and a radio-frequency electric field in an STM.

**Introduction**

Atomic and molecular spins on surfaces can provide a solid state qubit platform (*1,2*) that is unique in allowing bottom-up design capabilities. The coherent manipulations of single spins in a scanning tunneling microscope (STM) equipped with pulsed electron spin resonance (ESR) opened a way to utilize individual atomic (*3*) and molecular (*4*) spins on surfaces as feasible qubits for quantum information and computation. To date, magnetic resonance in STM (*5–9*) has relied on the electric-field driven modulation of the atomic scale magnetic interaction between the spin on the surface and the magnetic STM tip (*10–12*). A coherence time of a few tens of ns has been reported for the spins in the tunnel junction (*3,4,8*). This coherence is mainly limited by scattering of the spin states by tunneling electrons (*4,13*) and by the fluctuating magnetic field due to tip vibrations, limiting the number of available quantum gate operations within the coherence time. Moreover, such a 'tip-field-driven ESR' allows only one spin in the tunnel junction to be coherently controlled at a time, which has so far prohibited quantum manipulation of multiple spin qubits in STM. Therefore, it is necessary to develop a new mechanism to drive ESR of a spin in STM, free from strong coupling to the tip's magnetic field. Such a scheme has recently been realized with the demonstration of coherent control of spins that are outside the tunnel junction (*14*). Here, we show an elaborated study on the mechanism that enables the coherent control of such remote spins.

In this work, we present continuous wave (CW) and pulsed ESR of hydrogenated Ti atoms (spin $S = 1/2$) adsorbed on a thin MgO film (*15*), with Fe atoms located in close proximity, in an ultra-high vacuum STM operating at 1.2 K. Pulsed ESR of Ti atoms in such atomic pairs showed large Rabi rates, up to $\Omega/2\pi \approx 25$ MHz, even when the tip-Ti distance was large, which are comparable with the rates when driven by the interaction with the tip magnetic moment (*3,4*). CW-ESR on the Ti-Fe pair with an interatomic separation of 0.6–0.8 nm in a vector magnetic field revealed that the spin-spin interaction ranges up to 20 GHz and the driving strength is tunable as well. Combined with a model study, we show that such a driving field stems from the Fe atom's magnetic field gradient. The inhomogeneous local magnetic field provided by such a single-atom magnet is thus able to replace the role of the magnetic tip in coherent driving of a qubit in an ESR-STM (*6,9–12,16*).

**Results**

Our ESR-STM experiment on a pair of Ti and Fe atoms, hereafter referred to as a 'Ti-Fe pair', is illustrated in Fig. 1A. Both the tip and the Fe are coupled with the Ti spin via exchange interactions, $J_{\text{Ti,tip}}$ and $J_{\text{Ti,Fe}}$, respectively. The radio-frequency (RF) voltage $V_{\text{RF}}$ applied between tip and sample drives a spin resonance of the Ti, whose influence on the spin population is measured by the spin-polarized tunneling current (*5*) (see Materials and Methods). By utilizing atom manipulation techniques (*17*), we constructed such Ti-Fe pairs of three different separations (Figs. 1B-D) and performed ESR on the Ti atom of each pair as shown in Fig. 1E.

The ESR spectrum from the pair with the largest separation shown in Fig. 1B (black curve) exhibited only one resonance peak, determined by the Zeeman energy of an isolated Ti (*6,9*), which indicates a vanishingly small interaction with Fe. In contrast, Ti-Fe pairs of smaller separations (Figs. 1C and D) showed a clear splitting ($\Delta f$) of the resonance peak (red and blue curves in Fig. 1E), indicating a sizable magnetic interaction between Ti and Fe in the pair. This splitting was largest for the pair with the smaller separation (Fig. S1). On the MgO surface, the Fe spin fluctuates between spin up (⇑: $m_{\text{Fe}}$ = +2) and down (⇓: $m_{\text{Fe}}$ = –2) states, and this switching is slow enough that the peaks in the Ti spectrum are well resolved (Fig. S4) (*18*). This quasi-static behavior of Fe is consistent with isolated Fe atoms on MgO having a large out-of-plane uniaxial magnetic anisotropy energy (MAE) (*19*) and a spin relaxation time of a few tens of microseconds at the magnetic fields used here (*20,21*). As the Ti and Fe were positioned closer together, we observed significant shifts in the center frequency $f_c$ of resonance peak pairs by up to a few GHz. This shift arises from the change of the Ti spin direction due to the Ti-Fe interaction, which increases the Ti Zeeman energy due to the anisotropy of its g-factor (*9*) (Fig. S6).

Similar to the role of the exchange field from the Ti-tip interaction $J_{\text{Ti,tip}}$ in ESR of an isolated single spin (*10–12*), the Ti-Fe interaction $J_{\text{Ti,Fe}}$ can also effectively generate an inhomogeneous magnetic field at the location of Ti, which is able to drive the spin resonance of Ti. To demonstrate that the Ti-Fe interaction can indeed drive ESR of Ti, we performed pulsed ESR on the Ti atoms in each of the three Ti-Fe pairs shown in Figs. 1B-D and measured their Rabi rates ($\Omega$) as a function of tunnel conductance ($\sigma_{\text{tun}}$) by varying the tip-Ti distance (Fig. 2 and Fig. S7). At a large tunnel conductance, the measured Rabi rates were nearly independent of the Ti-Fe distance showing that driving was dominated by the Ti-tip interaction (Fig. 2A). In contrast, at small tunnel conductance, we found a strong dependence of the Rabi rate on the Ti-Fe distance (Fig. 2B). The Rabi rate for the largest Ti-Fe distance (effectively an isolated Ti atom) showed a linear decrease for a decreasing tunnel conductance (gray in Fig. 2C) that extrapolates to $\Omega \approx 0$ at zero tunnel conductance, which implies a vanishing driving field at a very large tip-atom distance. This linear behavior indicates an exponential dependence of the driving field on the tip height, a fingerprint for the exponential nature of the driving field contributed by the Ti-tip exchange interaction as previously observed (*11*).

In contrast, the Rabi rate of Ti positioned at sub-nanometer distances to Fe (Figs. 1C and D) remained finite, approaching a limiting value $\Omega_0$ (Fig. 2C). This strongly suggests that the ESR of Ti was driven effectively by the presence of the Fe spin at a tip height where the driving magnetic field from the tip was negligible. The driving RF electric field due the presence of the tip remains present at all tip heights tested, and decays only slowly (with inverse distance) as the tip is withdrawn from the Ti. Furthermore, we found that the zero conductance Rabi rate measured on the pair with a separation of 0.72 nm was a factor of ~2 smaller than that from the pair with 0.59 nm. The Rabi rate saturated at $\Omega_0/2\pi = 25.0 \pm 6.4$ MHz ) at $V_{RF} = 100$ mV for the closer pair (Figs. S8–S10), which is sufficient to coherently invert the spin state in ~20 ns (*3,4*). We note that this Rabi rate is comparable to that for the isolated Ti for the closest tip-Ti distances tested (*3*). This characteristic dependence of the Rabi rates on the tunnel conductance demonstrates that a single atom magnet is indeed able to coherently drive ESR of a single atom spin nearby.

For a given Ti-Fe pair, we can further discriminate between the influence of the tip spin and that of the Fe spin on the ESR of Ti by examining the dependence of Rabi rate on the tip height. Contribution from the tip to the Rabi rate is tunable by the tip-Ti distance, while that from the Fe is essentially independent of the tip-Ti distance. Figure 3A shows an evolution of CW ESR spectra as a function of tunnel conductance obtained from the pair with a separation of 0.72 nm. Here we make use of the CW-ESR to explore cases where the Rabi rate is too small to measure directly in the time domain. The ESR peak height $I_{ESR}$ is closely related to the Rabi rate according to $I_{ESR} \propto \Omega^2 T_1 T_2/(1 + \Omega^2 T_1 T_2)$, where $T_1$ is the energy relaxation time and $T_2$ is the dephasing time (Supplementary Text 6).The $f_⇑$ peak (the peak due to Fe in the spin-up state) showed a monotonic decrease of its height with decreasing conductance, reflecting a decreasing contribution of the tip spin to the driving field as the tip was moved away from the Ti atom; however, the peak height approached a non-zero limiting value as the tunnel conductance approached zero. In contrast, the height of the $f_⇓$ peak vanished at an intermediate tunnel conductance of about 0.15 nS and reappeared as the tip was withdrawn further from the atom. This behavior differs markedly from the ESR peak heights of isolated Ti atoms (Fig. S11) or Ti-Ti pairs (*8,15,22*), where the heights decrease monotonically and vanish at a very low conductance.

The vanishing ESR peak at $f_⇓$ results from a cancellation of the two Rabi rates contributed from the tip and Fe spins at a critical tip height, as follows: ESR of Ti in a Ti-Fe pair is driven by a vector sum of two driving fields, one stemming from the interactions with the tip and the other with the Fe single atom magnet. This is illustrated in the three insets in Fig. 3B. When the two driving fields are antiparallel, they can cancel each other and result in the disappearance of the peak for $f_⇓$ at a critical tip height, here corresponding to a conductance of 0.15 nS. On the other hand, the opposite orientation of the Fe spin leads to a parallel alignment of the two, which explains the monotonic decrease but the non-vanishing peak height for $f_⇑$ as the tunnel conductance approaches zero. This crossover demonstrates the presence of two competing transverse RF driving fields. Quantitative modeling of these summed driving fields gives an excellent agreement with the measurement (Figs. 3B–D).

Our measurements have shown that the spin-spin interaction in a Ti-Fe pair is the origin of the finite Rabi rates when the Ti-tip interaction is negligible as well as the peak splitting in the ESR spectra. To characterize the Ti-Fe interaction in more detail, we measured the dependence of the peak splitting in the ESR spectra of the two pairs on the field angle (Figs. 4A, B and Fig. S12). An isotropic exchange type Ti-Fe interaction $J_{\text{Ti,Fe}} S_{\text{Ti}} \cdot S_{\text{Fe}}$, which depends only on the Fe and Ti spin directions, is controllable by the direction of the external magnetic field. The Fe spin orientation is quasi-static and always aligned perpendicular to the MgO surface (*5,19*). Figures 4A and B show the resonance peak frequencies and peak splitting, respectively, measured as a function of the angle between the magnetic field and the surface normal ($\theta_{\text{ext}}$).

The splitting in both pairs reaches a maximum when the field is normal to the surface ($\theta_{\text{ext}} = 0$) and vanishes for in-plane fields ($\theta_{\text{ext}} = +90°$ or $–90°$), confirming that the Fe spin points along the surface normal. The Fe spin generates an effective field $\boldsymbol{B}_{\text{Fe}}$ at the position of the Ti atom, as depicted by $\boldsymbol{B}_{\text{Fe},\Uparrow}$ and $\boldsymbol{B}_{\text{Fe},\Downarrow}$ in Fig. 4C, resulting in the net field $\boldsymbol{B}_{\text{net}} = \boldsymbol{B}_{\text{ext}} + \boldsymbol{B}_{\text{Fe}}$, along which the Ti spin aligns (Supplementary Text 3). The peak splitting ($\Delta f$) from the Ti-Fe interaction is expected to show a cosine-dependence on the angle between the two spins $\eta = \theta_{\text{ext}} - \alpha$, with $\alpha$ formed by $\boldsymbol{B}_{\text{Fe}}$. The orientation of the Fe spin normal to the surface leads to a peak splitting of $|\cos\eta|$, which is mirror-symmetric about an in-plane direction of $\boldsymbol{B}_{\text{ext}}$. This accurately describes the features in the field angle dependence of the splitting in Fig. 4B and suggests that the peak splitting indeed originates only from the Ti-Fe interaction. Fitting the data in Fig. 4B resulted in exchange couplings $J_{\text{Ti,Fe}} = 1.1 \pm 0.1$ and $6.8 \pm 0.5$ GHz for the pairs of 0.72 nm and 0.59 nm, respectively.

Our experimental findings clearly show that a single atom magnet can indeed provide an atomic-scale and reliable local transverse field, which can be used to coherently drive ESR of a nearby Ti spin with oscillating electric field. To enhance the driving strength, a straightforward strategy is positioning more single atom magnets in proximity. To demonstrate this, we constructed a Fe-Ti-Fe structure by adding one more Fe atom near a Ti-Fe pair of 0.72 nm, as shown in Fig. 5A, and performed ESR measurements on the Ti atom. The ESR spectrum from this spin complex showed four resonance peaks (Fig. 5B), corresponding to four spin orientations of two Fe spins, $|\Uparrow\Uparrow\rangle$, $|\Downarrow\Uparrow\rangle$, $|\Uparrow\Downarrow\rangle$, and $|\Downarrow\Downarrow\rangle$ (Fig. S5). To figure out the effect of two Fe atoms on the ESR of the Ti spin, we performed pulsed ESR on this complex at the resonance peak $f_3$ (Fig. S15). Figure 5C shows the Rabi rate as a function of the tunnel conductance (purple) as well as that measured on a Ti-Fe pair with the same tip (red). This comparison directly reveals an enhancement of the Fe-induced driving field by a factor of about two, indicating that our atomic-scale and bottom-up approach is plausible to tailor magnetic resonances in artificial atomic structures.

**Discussion**

To have a rigorous understanding on how the spin interactions influenced the ESR spectra, we set up a time-independent Hamiltonian for the spin configurations in this work as illustrated in Fig. 1A:

$$\widehat{H}_0 = -\mu_B g_{\text{Ti}} \boldsymbol{S}_{\text{Ti}} \cdot \boldsymbol{B}_{\text{ext}} + J_{\text{Ti,Fe}} \boldsymbol{S}_{\text{Ti}} \cdot \boldsymbol{S}_{\text{Fe}} + J_{\text{Ti,tip}} \boldsymbol{S}_{\text{Ti}} \cdot \boldsymbol{S}_{\text{tip}}, \qquad (1)$$

where $g_{Ti}$ is the g-factor of the Ti spin. According to the simplified models of the spin states of Fe (Fig. S4) and tip (Fig. S13) (9), we treat the Fe and the tip's spins as classical magnetic moments of spin expectation values $\langle S_{Fe} \rangle$ and $\langle S_{tip} \rangle$ pointing along the directions $\pm \hat{z}$ and $\pm \boldsymbol{n}_{tip}$, respectively. Here $\boldsymbol{n}_{tip} = (n_x, n_y, n_z)$ is determined by the uniaxial magnetic anisotropy of the tip magnetic moment. We obtained $\boldsymbol{n}_{tip}$ from field-angle ($\theta_{ext}$) dependence of ESR frequency, peak height, and Rabi rate measured on an isolated Ti atom (Fig. S13). Eqn. 1 can be reduced to a relatively simple form:

$$\widehat{H}_0 = \boldsymbol{S}_{Ti} \cdot \widetilde{\boldsymbol{B}}_0,$$

$$\widetilde{\boldsymbol{B}}_0 = \big(-g_x \mu_B B_x + J_{Ti,tip} \langle S_{tip} \rangle n_x, -g_y \mu_B B_y + J_{Ti,tip} \langle S_{tip} \rangle n_y, -g_z \mu_B B_z + J_{Ti,tip} \langle S_{tip} \rangle n_z + J_{Ti,Fe} \langle S_{Fe} \rangle\big), \qquad (2)$$

where $\widetilde{\boldsymbol{B}}_0$ is a generalized magnetic field with four different cases, one for each sign of $\langle S_{Fe} \rangle$ and $\langle S_{tip} \rangle$. Diagonalization of $\widehat{H}_0$ yields eigenstates, leading to four available ESR transitions of the Ti spin corresponding to four distinct combinations of the Fe and tip's spin states, $|\Uparrow\rangle_{Fe}|\Uparrow\rangle_{tip}$, $|\Downarrow\rangle_{Fe}|\Uparrow\rangle_{tip}$, $|\Uparrow\rangle_{Fe}|\Downarrow\rangle_{tip}$, and $|\Downarrow\rangle_{Fe}|\Downarrow\rangle_{tip}$. This dependence of the eigenstates on the tip spin is responsible for the periodic evolution of the resonance frequencies with the field angle by 180°, as observed from the pair of 0.72 nm (red markers in Fig. 4A; Fig. S13). An intuitive feature of $\widetilde{\boldsymbol{B}}_0$ is found in the case when the tip is far enough from Ti to ignore the Ti-tip interaction ($J_{Ti,tip} \ll J_{Ti,Fe}$), so that the splitting becomes symmetric about $\theta_{ext} = \pm 90°$. An example can be seen in the data from the pair of 0.59 nm (blue markers in Fig. 4), which were measured at a tunnel conductance of 0.05 nS ($V_{DC}$ = 200 mV, $I_{DC}$ = 10 pA) (Fig. S14).

Using the dependences of the eigenenergies of $\widehat{H}_0$ on the field-angle ($\theta_{ext}$) and tuning $J_{Ti,Fe}$, we fit the resonance frequencies and peak splitting (Figs. 4A and B), which excellently reproduced the detailed experimental features (solid curves in Figs. 4A and B). The asymmetry in the resonance frequencies about $\theta_{ext} = +90°$ and $-90°$ observed from the pair of 0.72 nm is revealed to stem from a sizable contribution of the tip spin at the relatively small tip-Ti distance (red markers; $V_{DC}$ = 30 mV, $I_{DC}$ = 20 pA).

By considering both Ti-tip and Ti-Fe interactions, which contribute to the coherent RF driving fields $\boldsymbol{B}_{1,tip}$ and $\boldsymbol{B}_{1,Fe}$, respectively, the total driving field becomes $\boldsymbol{B}_1 = \boldsymbol{B}_{1,tip} + \boldsymbol{B}_{1,Fe}$, as illustrated in Fig. 3C. In ESR of a Ti spin ($S$ = 1/2), only the components of $\boldsymbol{B}_1$ perpendicular to the total static field ($B_0$) contribute to the transverse time-varying field $B_{1\perp}$, such that the Rabi rate can be described by

$$\Omega = g\mu_B m_S |B_{1,\text{tip}\perp} + B_{1,\text{Fe}\perp}|/\hbar, \qquad (3)$$

with the magnetic quantum number of the Ti spin ($m_S$ = 1/2). For some tips, such as the one shown in Fig. 3, the two driving fields are either parallel ($\phi = 0$) or antiparallel ($\phi = 180°$), corresponding to the $|\Uparrow\rangle$ or $|\Downarrow\rangle$ states of the Fe, respectively (insets of Fig. 3B). For $\phi = 180°$, the net driving field ($B_{1\perp}$) goes to zero at a critical Ti-tip distance so that the ESR peak vanishes, and then it reappears with the opposite sign, resulting in the opposite Rabi rotation of the Bloch vector with a reappearance of the ESR signal (Fig. 3D). For this special case of parallel driving fields, the Rabi rate of the Ti spin in Eqn. (3) can be rewritten as a simple sum of two contributions $\Omega_\pm = |\Omega_{\text{tip}} \pm \Omega_{\text{Fe}}|$, where the + (−) sign represents the parallel (antiparallel) case. This model using the $\Omega_\pm$ excellently reproduced the tunnel conductance dependence of the ESR peak heights (Supplementary Text 6; solid curves in Fig. 3B). The extrapolation of the peak heights to zero conductance yields finite intercepts of a peak height ratio ~0.64, which is largely determined by the thermal occupations of the two Fe states at the measurement temperature of 1.2 K.

In the following, we show a quantification of the Fe-induced driving field according to a model for ESR of a single spin on surface, where the RF piezoelectric response $\Delta Z_1(t)$ of the atom converts the RF electric field $E(t)$ into a time-varying magnetic field $B_1(t)$ at the position of the atom (*10,11*). Similar to ESR contributed from the Ti-tip interaction, the Ti-Fe interaction can play the same role when coupled with the RF voltage $V_{\text{RF}}(t)$ applied between the tip and substrate (Supplementary Text 7). The Rabi frequency contributed from the Fe takes a form of

$$\Omega_{\text{Fe}} = (\partial J_{\text{Ti,Fe}}/\partial z) \cdot \Delta Z_1 \cdot \sin\theta_{\text{Ti}}, \qquad (4)$$

where $\theta_{\text{Ti}}$ is the polar angle of the Ti spin (Fig. S17). We assume an isotropic Ti-Fe interaction with a decay length $d_{\text{ex}}$ = 86 pm (Supplementary Text 1) and take an adiabatic approximation to the RF displacement $\Delta Z_1(t)$ in our frequency range of 20–30 GHz due to the bonding strengths of ~ a few THz for both atoms to MgO. Here piezoelectric responses of both Ti and Fe atoms should be taken account, such that the displacements of Ti by both static and RF electric fields are measured relative to those of Fe. Using $J_{\text{Ti,Fe}}$ = 6.68 GHz for the pair of 0.59 nm as obtained in this work (Fig. 4B, Fig. S1), Eqn. 4 yields $\Delta Z_1/V_{RF} \approx 0.20$ pm/mV as the RF displacement for the Rabi rate $\Omega_{Fe}/2\pi \approx 25$ MHz as measured when the tip's contribution was negligible (Fig. 2C; $V_{DC}$ = 100 mV, $I_{DC}$ = 10 pA). This is comparable to the value (~0.29 pm/mV) derived for ESR of an isolated Ti atom using the same model in a previous report (*11*). By considering piezoelectric motions of the two atoms of similar magnitudes but in opposite phases, we obtain $\Delta Z_{1,Ti}/V_{RF} \approx \Delta Z_{1,Fe}/V_{RF} \sim 0.10$ pm/mV. However, this is larger by two orders of magnitude than a theoretical estimation reported earlier (*6*) when considering only the stretching of the atom-MgO bonds as the sources of $\Delta Z_1$. Further theoretical studies are desired for insights into other significant contributions to $\Delta Z_1$, such as the piezoelectric motion of the MgO layer and non-linear piezoelectric responses of the atoms (*6*).

Utilizing a single atom magnet for ESR driving liberates on-surface spins, which have been limited to a single qubit in the conventional ESR-STM configuration. Furthermore, this approach enables the separation of driving and detection of ESR, providing longer relaxation and quantum coherence times (*14*). Importantly, it enables atomic and molecular spins on surfaces as a platform of multi qubit operations. In this work, we investigated the electron spin resonance of a single Ti spin on MgO with Fe atoms in close proximity, unraveling the underlying driving mechanism of ESR in such a spin, exchange-coupled to a single-atom magnet. Together with atom manipulation technique and the availability of even better single-atom magnets (*23–25*), our work sheds light on an atomically precise design of on-surface multi-spin qubit structures with long relaxation and coherence times.

**Materials and Methods**

An atomically clean Ag(100) substrate was prepared by alternating Ar ion sputtering and annealing cycles. MgO films were grown on the Ag substrate at 580 K by evaporating Mg in an $O_2$ atmosphere of $1.1\times10^{-6}$ Torr. Then, Fe and Ti atoms were deposited on the pre-cooled MgO surface. All measurements were performed on Ti atoms bound to a bridge site of the MgO surface (that is, in the middle of two oxygen sites). Before measurements, the STM tip which was made of Pt/It wire was poked into the Ag(001) surface until satisfactory topographic and spectroscopic features were observed on atoms on MgO, after which Fe atoms were picked up by the STM tip from MgO (by applying a DC voltage pulse of 0.3 V) to create a spin-polarized tip. The tip's spin polarization was calibrated with the asymmetry around zero bias in the d$I$/d$V$ spectra of Ti on MgO.

Measurements were performed in two ultrahigh-vacuum ($< 10^{-10}$ mbar) scanning tunneling microscopes (STMs). The data presented in Figs. 1, 2, 3, 5 were measured in a home-built $^3$He-cooled STM at $T$ = 1.2 K equipped with a single-axis superconducting magnet. The data presented in Fig. 4 and Figs. S12–S14 were measured in a commercial $^3$He-cooled STM (Unisoku, USM1300) at $T$ = 0.4 K equipped with two-axis superconducting magnets. High-frequency transmission cables were installed on the STM systems as described in detail elsewhere (*20,26*). Continuous wave electron spin resonance (ESR) spectra were acquired by sweeping the frequency of an RF voltage $V_{\mathrm{RF}}$ generated by an RF generator (Agilent E8257D) across the tunneling junction and monitoring changes in the tunneling current. For pulsed ESR, the RF generator was gated by the pulse outputs programed in arbitrary waveform generators (Tektronix 7122C for 1K-system; Tektronix 5000 for 0.4K-system). The output signal of the RF generator was combined with a DC bias voltage through a bias tee (SigaTek, SB15D2). In both CW- and pulsed-ESR measurements, the RF signals were chopped at 95 Hz and sent to a lock-in amplifier (Stanford Research Systems SR860) and recorded by a DAQ-Box (National Instruments 6363). The bias voltage $V_{\mathrm{DC}}$ refers to the sample voltage relative to the tip. The STM constant-current feedback loop was set to a low gain during the measurements.

**References:**

(1) A. J. Heinrich, W. D. Oliver, L. Vandersypen, A. Ardavan, R. Sessoli, D. Loss, A. B. Jayich, J. Fernandez-Rossier, A. Laucht, A. Morello, Quantum-coherent Nanoscience. *Nature Nanotech.* **16**, 1318–1329 (2021).

(2) S. Thiele, F. Balestro, R. Ballou, S. Klyatskaya, M. Ruben, W. Wernsdorfer, Electrically driven nuclear spin resonance in single-molecule magnets. *Science* **344**, 1135-1138 (2014).

(3) K. Yang, W. Paul, S. H. Phark, P. Willke, Y. Bae, T. Choi, T. Esat, A. Ardavan, A. J. Heinrich, C. P. Lutz, Coherent spin manipulation of individual atoms on a surface. *Science* **366**, 509-512 (2019).

(4) P. Willke, T. Bilgeri, X. Zhang, Y. Wang, C. Wolf, H. Aubin, A. J. Heinrich, T. Choi, Coherent Spin Control of Single Molecules on a Surface. *ACS Nano* **15**, 17959–17965 (2021).

(5) S. Baumann, W. Paul, T. Choi, C. P. Lutz, A. Ardavan, A. J. Heinrich, Electron paramagnetic resonance of individual atoms on a surface. *Science* **350**, 417-420 (2015).

(6) T. Seifert, S. Kovarik, D. M. Juraschek, N. A. Spaldin, P. Gambardella, S. Stepanow, Longitudinal and transverse electron paramagnetic resonance in a scanning tunneling microscope, *Sci. Adv.* **6**, eabc5511 (2020).

(7) M. Steinbrecher, W. M. J. van Weerdenburg, E. F. Walraven, N. P. E. van Mullekom, J. W. Gerritsen, F. D. Natterer, D. I. Badrtdinov, A. N. Rudenko, V. V. Mazurenko, M. I. Katsnelson, Ad van der Avoird, G. C. Groenenboom, A. A. Khajetoorians, Quantifying the interplay between fine structure and geometry of an individual molecule on a surface. *Phys. Rev. B* **103**, 155405 (2021).

(8) L. M. Veldman, L. Farinacci, R. Rejali, R. Broekhoven, J. Gobeil, D. Coffey, M. Ternes, A. F. Otte, Free coherent evolution of a coupled atomic spin system initialized by electron scattering. *Science* **372**, 964-968 (2021).

(9) J. K. Kim, W. Jang, H. T. Bui, D. Choi, C. Wolf, F. Delgado, D. Krylov, S. Lee, S. Yoon, C. P. Lutz, A. J. Heinrich, Y. Bae, Spin Resonance Amplitude and Frequency of a Single Atom on a Surface in a Vector Magnetic Field. *Phys. Rev. B* **104**, 174408 (2021).

(10) J. L. Lado, A. Ferrón, J. Fernández-Rossier, Exchange mechanism for electron paramagnetic resonance of individual adatoms. *Phys. Rev. B* **96**, 205420 (2017).

(11) K. Yang, W. Paul, F. D. Natterer, J. L. Lado, Y. Bae, P. Willke, T. Choi, A. Ferrón, J. F-Rossier, A. J. Heinrich, C. P. Lutz, Tuning the Exchange Bias on a Single Atom from 1 mT to 10 T. *Phys. Rev. Lett.* **122**, 227203 (2019).

(12) J. Reina Gálvez, C. Wolf, F. Delgado, N. Lorente, Cotunneling mechanism for all-electrical electron spin resonance of single adsorbed atoms. *Phy. Rev. B* **100**, 035411 (2019).

(13) P. Willke, W. Paul, F. D. Natterer, K. Yang, Y. Bae, T. Choi, J. Fernández-Rossier, A. J. Heinrich, C. P. Lutz, Probing quantum coherence in single-atom electron spin resonance, *Sci. Adv*. **4**, eaaq1543 (2018).

(14) Y. Wang, Y. Chen, H. T. Bui, C. Wolf, M. Haze, C. Mier, J. Kim, D.-J. Choi, C. P. Lutz, Y. Bae, A. J. Heinrich, S. Phark, An electron-spin qubit platform crafted atom-by-atom on a surface. *arXiv*:2108.09880 (2022).

(15) K. Yang, Y. Bae, W. Paul, F. D. Natterer, P. Willke, J. L. Lado, A. Ferrón, T. Choi, J. Fernández-Rossier, A. J. Heinrich, C. P. Lutz, Engineering the Eigenstates of Coupled Spin-1/2 Atoms on a Surface. *Phys. Rev. Lett.* **119**, 227206 (2017).

(16) A. Ferrón, S. A. Rodríguez, S. S. Gómez, J. L. Lado, J. Fernández-Rossier, Single Single spin resonance driven by electric modulation of the g-factor anisotropy. *Phys. Rev. Res.* **1**, 033185 (2019).

(17) D. M. Eigler, E. K. Schweizer, Positioning single atoms with a scanning tunneling microscope. *Nature* **344**, 524–526 (1990).

(18) T. Choi, W. Paul, S. Rolf-Pissarczyk, A. J. Macdonald, F. D. Natterer, K. Yang, P. Willke, C. P. Lutz, A. J. Heinrich, Atomic-scale sensing of the magnetic dipolar field from single atoms. *Nat. Nanotech*. **12**, 420–424 (2017).

(19) S. Baumann, F. Donati, S. Stepanow, S. Rusponi, W. Paul, S. Gangopadhyay, I. G. Rau, G. E. Pacchioni, L. Gragnaniello, M. Pivetta, J. Dreiser, C. Piamonteze, C. P. Lutz, R. M. Macfarlane, B. A. Jones, P. Gambardella, A. J. Heinrich, and H. Brune, Origin of Perpendicular Magnetic Anisotropy and Large Orbital Moment in Fe Atoms on MgO. *Phys. Rev. Lett.* **115**, 237202 (2015).

(20) W. Paul, K. Yang, S. Baumann, N. Romming, T. Choi, C. P. Lutz, A. J. Heinrich, Control of the millisecond spin lifetime of an electrically probed atom. *Nat. Phys*. **13**, 403–407 (2017).

(21) P. Willke, A. Singha, X. Zhang, T. Esat, C. P. Lutz, A. J. Heinrich, T. Choi, Tuning single-atom electron spin resonance in a vector-magnetic field. *Nano Lett.* **19**, 8201–8206 (2019).

(22) Y. Bae, K. Yang, P. Willke, T. Choi, A. J. Heinrich, C. P. Lutz, Enhanced quantum coherence in exchange coupled spins via singlet-triplet transitions. *Sci. Adv.* **4**, eaau4159 (2018).

(23) F. Donati, S. Rusponi, S. Stepanow, C. Wäckerlin, A. Singha, L. Persichetti, R. Baltic, K. Diller, F. Patthey, E. Fernandes, J. Dreiser, Ž. Šljivančanin, K. Kummer, C. Nistor, P. Gambardella, H. Brune, Magnetic remanence in single atoms. *Science* **352**, 318-321 (2016).

(24) F. D. Natterer, K. Yang, W. Paul, P. Willke, T. Choi, T. Greber, A. J. Heinrich, C. P. Lutz, Reading and writing single-atom magnets. *Nature* **543**, 226-228 (2017).

(25) A. Singha, P. Willke, T. Bilgeri, X. Zhang, H. Brune, F. Donati, A. J. Heinrich, T. Choi, Engineering atomic-scale magnetic fields by dysprosium single atom magnets. *Nat. Comm.* **12**, 1-6 (2021).

(26) J. Hwang, D. Krylov, R. Elbertse, S. Yoon, T. Ahn, J. Oh, L. Fang, W. Jang, F. H. Cho, A. J. Heinrich, Y. Bae, Development of a scanning tunneling microscope for variable temperature electron spin resonance. *Rev. Sci. Instrum.* **93**, 093703 (2022).

(27) F. Delgado, J. Fernández-Rossier, Spin decoherence of magnetic atoms on surfaces. *Prog. Surf. Sci*. **92**, 40–82 (2017).

(28) M. A. Subramanian, R. D. Shannon, B. H. T. Chai, M. M. Abraham, M. C. Wintersgill, Dielectric constants of BeO, MgO, and CaO using the two-Terminal method, *Phys. Chem. Minerals* **16**, 741-746 (1989).

(29) P Giannozzi et al. Advanced capabilities for materials modelling with Quantum ESPRESSO. *J. Phys.*: *Condens. Matter* **29**, 465901 (2017).

(30) Andrea Dal Corso, Pseudopotentials periodic table: From H to Pu. *Computational Materials Science* **95**, 337-350 (2014).

(31) S. Grimme, A. Hansen, J. G. Brandenburg, C. Bannwarth, Dispersion-Corrected Mean-Field Electronic Structure Methods. *Chem. Rev*. **116**, 5105–5154 (2016).

**Acknowledgments**

We thank Harald Brune for fruitful discussions and comments on the manuscript.

**Funding:**

U.S. Department of Navy under award N00014-21-1-2467 issued by the Office of Naval Research (CPL).
Institute for Basic Science grant IBS-R027-D1 (SP, HTB, YW, JR-G, CW, AJH).
FCT grant PTDC/FIS-MAC/2045/2021 (JFR)
Swiss National Foundation Sinergia grant Pimag (JFR)
Generalitat Valenciana grant Prometeo2021/017 and MFA/2022/045 (JFR)
MICIIN-Spain grant PID2019-109539GB-C41 (JFR)

**Author contributions:**

Conceptualization: CPL, SP, AJH
Experiments and data analysis: SH, HTB, YW, KY, CPL
Modeling: AF, JFR
Numerical simulations and data fitting: JR-G, CW, HTB
Writing: all authors

**Competing interests:** All authors declare no competing interests.

**Data and materials availability:** All data are available in the main text or the supplementary materials.

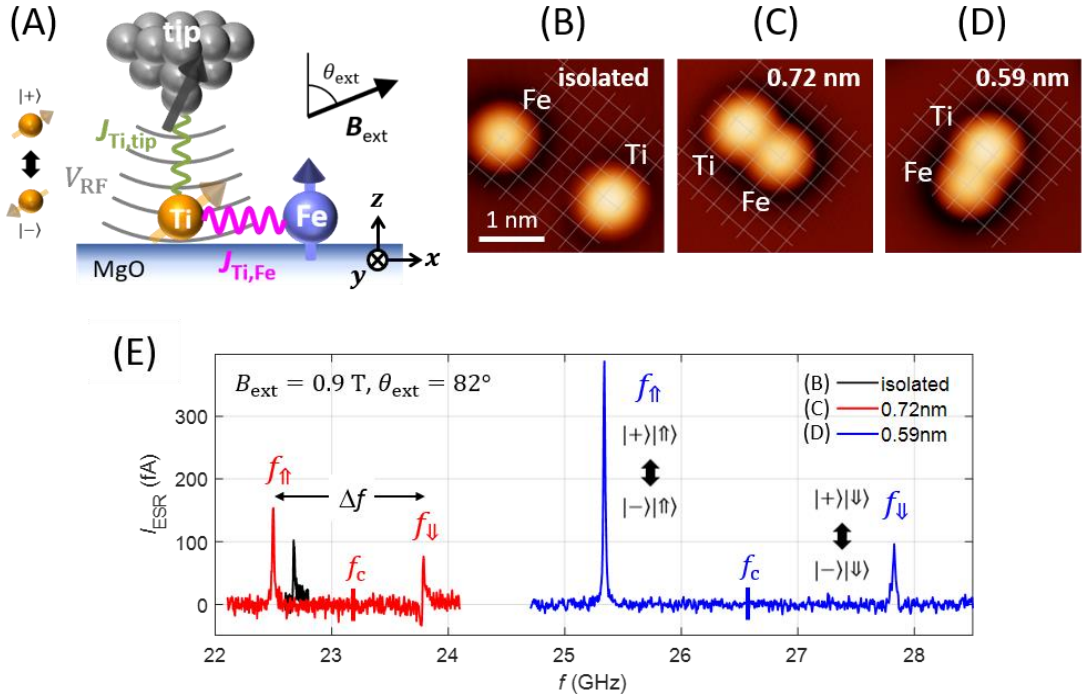

**Fig. 1. Electron spin resonance (ESR) in Ti-Fe pairs.** (**A**) Schematic showing a Ti spin ($S = 1/2$) coupled with Fe on MgO in electron spin resonance (ESR). $J_{\text{Ti,Fe}}$ and $J_{\text{Ti,tip}}$ are Ti-Fe and Ti-tip spin-spin interactions, which drive ESR of the Ti spin when coupled with the RF voltage $V_{\text{RF}}$ applied to the tunnel junction. (**B-D**) STM images of Ti atoms with Fe positioned at distances of 1.9 nm, 0.72 nm, and 0.59nm away from Ti, respectively ($V_{\text{DC}}$ = 50 mV, setpoint tunnel current $I_{\text{tun}}$ = 10 pA). (**E**) Continuous wave ESR spectra measured on Ti atoms in (B) (black), (C) (red), and (D) (blue) at $B_{\text{ext}}$ = 0.9 T with the polar angle $\theta_{\text{ext}}$ = 82º ($V_{\text{DC}}$ = 50 mV, $V_{\text{RF}}$ = 20 mV, $I_{\text{tun}}$ = 15 pA, $T$ = 1.2 K). The frequency $f_c$ indicates the center of the two peaks in each curve of the pairs.

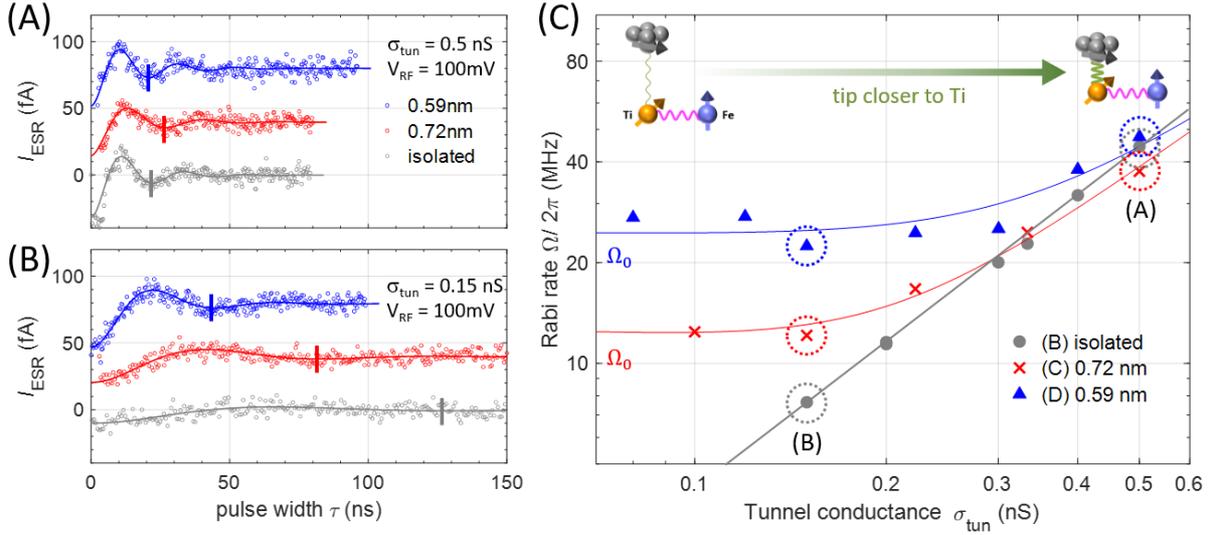

**Fig. 2. Influence of Fe on the Rabi rate.** (**A** and **B**) Rabi oscillations using pulsed ESR on Ti atoms shown in Figs. 1B–D at junction conductance of (A) 0.5 nS and (B) 0.15 nS, $V_{RF}$ = 100 mV. Oscillation period of each curve is marked by the vertical bar. The curves are successively shifted vertically by 40 fA for clarity. (**C**) Rabi rate $\Omega$ extracted from Rabi oscillation measurements of isolated Ti (gray circles), Ti-Fe of 0.72 nm (red crosses), and Ti-Fe of 0.59 nm (blue triangles) ($I_{tun}$ = 10 pA, $V_{RF}$ = 100 mV, $T$ = 1.2 K, $B_{ext}$ = 0.9 T, $\theta_{ext}$ = 82°). The dotted circles denote the data points corresponding to the curves in (A) and (B) with the same color codes. Two insets illustrate the magnetic interactions, $J_{Ti,Fe}$ and $J_{Ti,tip}$, for small (left) and large (right) tunnel conductance regimes, respectively. Solid curves are fits using the model described in the Supplementary Text 4, resulting in zero conductance Rabi rates $\Omega_0/2\pi$ of $27 \pm 2$ and $14 \pm 2$ MHz for the pairs of 0.59 and 0.72 nm, respectively.

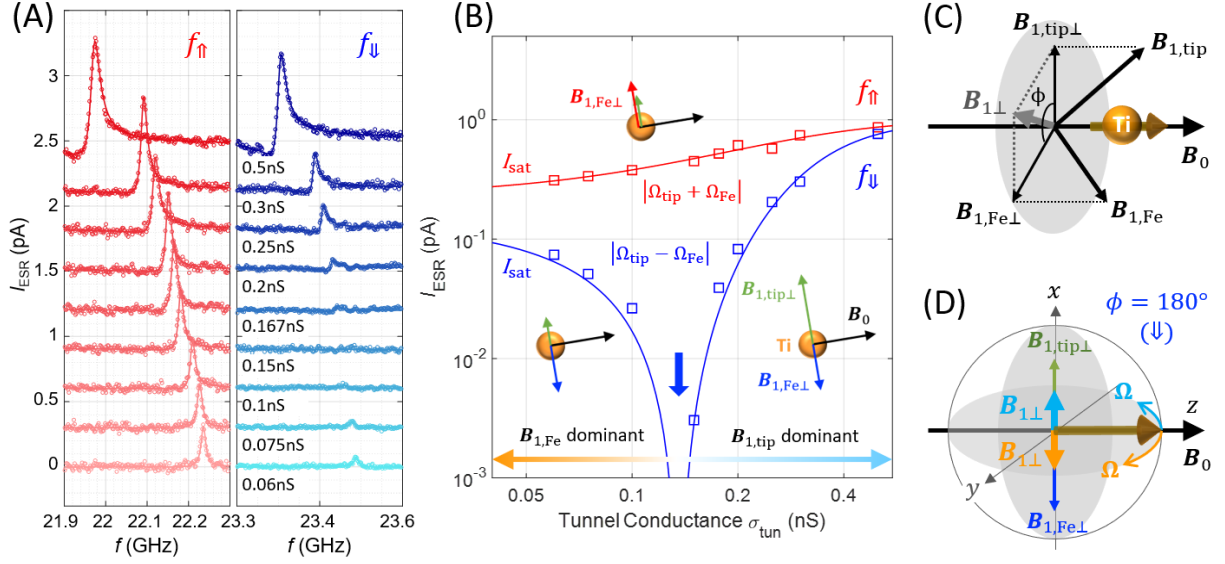

**Fig. 3. Crossover from tip-driven to Fe-driven ESR.** (**A**) CW-ESR spectra measured on the Ti-Fe of 0.72 nm at frequency ranges across the peaks $f_⇑$ (red) and $f_⇓$ (blue) for a varying tunnel conductance ($V_{DC}$ = 50 mV, $I_{tun}$ = 15 pA, $V_{RF}$ = 20 mV, $T$ = 1.2 K, $B_{ext}$ = 0.9 T, $θ_{ext}$ = 82°). The solid curves are asymmetric Lorentzian fits. The curves are successively shifted vertically by 0.3 pA for clarity. (**B**) Peak height $I_{ESR}$ vs. tunnel conductance extracted from CW-ESR spectra in (A). Fits according to the discussion in the main text are overlaid (solid curves; see DISCUSSION section). Extrapolation of the fit curves to zero conductance gives intercepts $I_{sat}$ = 0.226 pA ($f_⇑$) and 0.142 pA ($f_⇓$). The three insets illustrate the vectorial relationships between the ESR driving fields contributed from the Fe (red or blue) and the tip (light green). (**C**) Vectorial relationship of driving fields from Fe ($\boldsymbol{B}_{1,Fe}$) and tip ($\boldsymbol{B}_{1,tip}$). $\boldsymbol{B}_0$ is the total static field at the Ti position, composed of external ($\boldsymbol{B}_{ext}$), tip-induced field ($\boldsymbol{B}_{tip}$), and Fe-induced ($\boldsymbol{B}_{Fe}$) fields. $\boldsymbol{B}_{1,Fe⊥}$ and $\boldsymbol{B}_{1,tip⊥}$ denote projections of $\boldsymbol{B}_{1,Fe}$ and $\boldsymbol{B}_{1,tip}$, respectively, to a plane perpendicular to the total static field $\boldsymbol{B}_0$. (**D**) A schematic of Bloch sphere in a condition that $\boldsymbol{B}_{1,Fe⊥}$ and $\boldsymbol{B}_{1,tip⊥}$ are antiparallel ($ϕ$ = 180°) and showing resultant Rabi rotations of the Bloch vector of the Ti spin (brown thick arrow). The net driving field $\boldsymbol{B}_{1⊥}$ changes its direction depending on the magnitude of the tip-induced driving field $\boldsymbol{B}_{1,tip⊥}$, leading to corresponding change in the Rabi rotation.

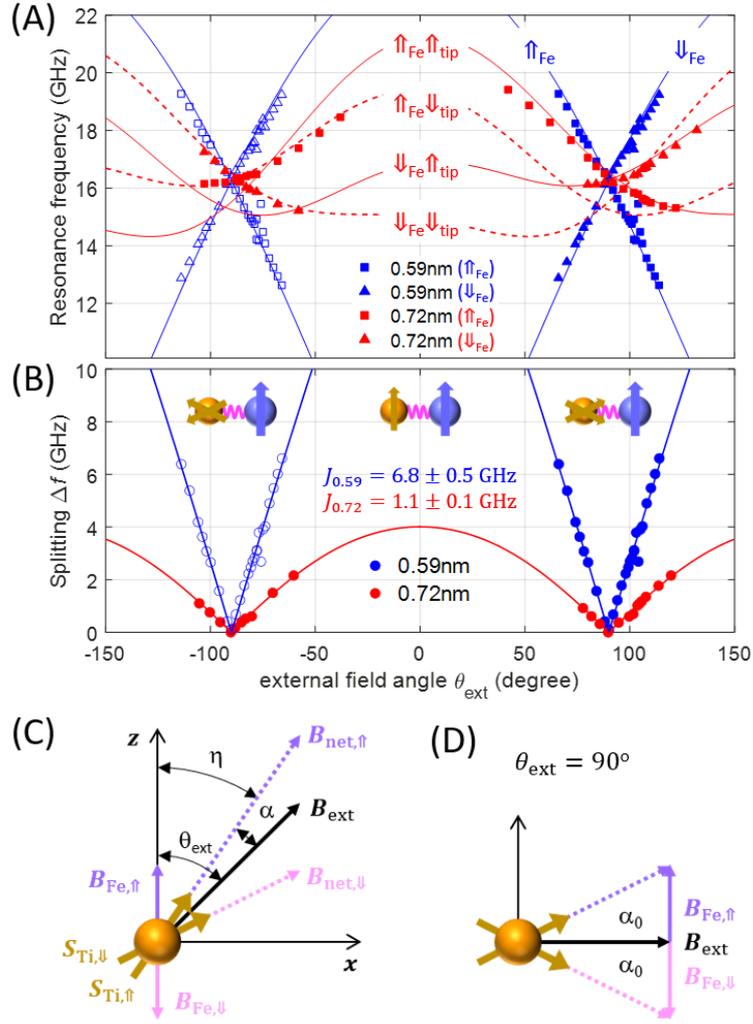

**Fig. 4. Angle dependence of ESR frequencies on the applied magnetic field.** (**A** and **B**) ESR resonance peak frequencies (A) and splitting $\Delta f$ (B) in CW-ESR spectra measured from Ti-Fe pairs of 0.72 nm (red) and 0.59 nm (blue) at a varying polar angle ($\theta_{ext}$) of $\boldsymbol{B}_{ext}$. Solid curves are fits using the model described in the equations (5) and (6). Red solid (dashed) curves correspond to the up (down) state of the tip spin. ($V_{DC}$ = 200 mV, $I_{tun}$ = 10 pA, $V_{RF}$ = 30 mV for the 0.59 nm pair; $V_{DC}$ = 30 mV, $I_{tun}$ = 20 pA, $V_{RF}$ = 30 mV for the 0.72 nm pair; $T$ = 0.4 K, $B_{ext}$ = 0.6 T). (**C** and **D**) Schemes of net magnetic field $\boldsymbol{B}_{net}$ at the Ti position, composed of the external field $\boldsymbol{B}_{ext}$ and Fe-induced field $\boldsymbol{B}_{Fe}$ in the plane of $\boldsymbol{B}_{ext}$ vector with an arbitrary (C) and 90° (D) field angle $\theta_{ext}$. $\alpha$ and $\alpha_0$ denote the angles added by the Fe-induced field.

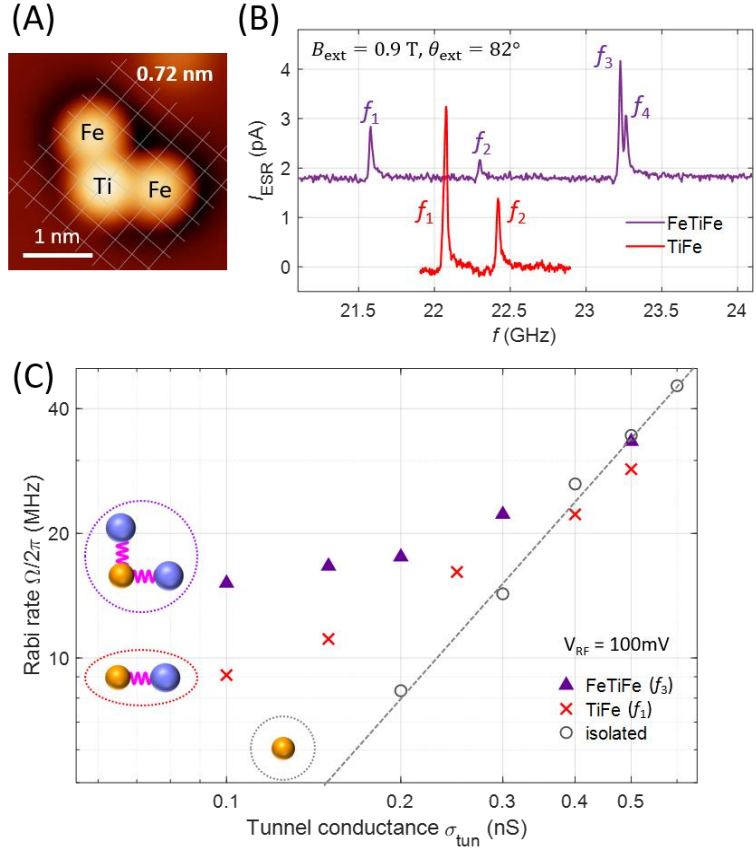

**Fig. 5. ESR of Ti using two Fe atoms.** (**A**) A Fe-Ti-Fe spin complex with Ti-Fe separations of 0.72nm. (**B**) Continuous wave ESR spectra measured on Ti of the complex in (A) (purple) and a Ti-Fe pair of 0.72nm (red) ($V_{DC}$ = 50 mV, $I_{tun}$ = 15 pA, $V_{RF}$ = 100 mV, $T$ = 1.2 K, $B_{ext}$ = 0.9 T, $\theta_{ext}$ = 82°). (**C**) Rabi rates obtained from pulsed ESR of the complex (purple triangles) and TiFe pair of 0.72nm (red crosses), and an isolated Ti (gray open circles) at a varying tunnel conductance. The insets schematically depict the Ti-Fe interactions for three groups of Rabi rates ($V_{RF}$ = 100 mV, $T$ = 1.2 K, $B_{ext}$ = 0.9 T, $\theta_{ext}$ = 82°).

# Supplementary Materials for

**Electric-field-driven spin resonance by exchange coupling to a single-atom magnet**

Soo-hyon Phark *et al.*

*Corresponding authors. Email: heinrich.andreas@qns.science, cplutz@us.ibm.com

**This PDF file includes:**

Supplementary Text
Figs. S1 to S18

**Supplementary Text**

**Section 1:** ESR peak splitting measured on Ti-Fe pairs

1.1. Dependence on Ti-Fe separation

As discussed in the main text (Fig. 1E), the ESR peak splitting measured on a Ti-Fe pair monotonically increased for a decreasing separation between two atoms. Figures S1A and B show ESR spectra measured on Ti-Fe pairs of five different separations and peak splitting with respect to the Ti-Fe separation. Assuming an isotropic exchange interaction, we applied an exponential dependence of splitting ($\Delta f$) on Ti-Fe separation ($r_{\text{Ti,Fe}}$) (*15,22*)

$$\Delta f(r_{\text{Ti,Fe}}) = J_0 \exp[-(r_{\text{Ti,Fe}} - r_0)/d_{\text{ex}}], \tag{S1}$$

where $J_0$ and $d_{\text{ex}}$ are the exchange coupling energy at $r_{\text{Ti,Fe}} = r_0$ and the decay length of the Ti-Fe interaction. Fitting the data in Fig. S1B to the equation (S1) with $r_0 = 0.59$ nm resulted in $d_{\text{ex}} = 86 \pm 27$ pm and $J_0 = 3.22 \pm 0.45$ GHz (red curve).

1.2. Dependence on pair orientations

We note that the pairs with a separation of 0.72 nm showed a mean splitting of 0.9 GHz but with a large dispersion of about 0.5 GHz (Fig. S1). Due to the out-of-plane MAE of the Fe spin, a possible contribution from dipole-dipole interaction should have the same form as that from exchange interaction, leading to a total spin-spin interaction to be a simply combined shape $(J - D)\, \mathbf{S}_{\text{Ti}} \cdot \mathbf{S}_{\text{Fe}}$, where $D$ is the dipole-dipole coupling strength for a given Ti-Fe separation. Thus, we expect that the total spin-spin interaction should be isotropic for the azimuthal orientation of a pair. To survey this, we measured the ESR peak splitting on Ti-Fe pairs of 0.72 nm with different pair orientations as shown in Fig. S2. The splitting is anisotropic and doesn't show any regular dependence on the pair orientation. We speculate that this is from the difference in local environment of each pair and/or existence of defects nearby, unresolvable using normal STM microscopy and spectroscopy.

1.3. Dependence on tips

One possible origin of such a large dispersion of peak splitting observed from pairs of 0.72 nm is the anisotropic local field from the tip spin since it contributes an additional Zeeman energy to the atomic spins, which is comparable in size to the spin-spin interactions in the pair. We performed ESR on four pairs ('1' – '4') indicated in Fig. S2 with 21 different tips and show the measured splitting in Fig. S3. The mean splitting of the four pairs showed a large variation of 0.3–1.3 GHz as that shown in Fig. S2 for the same pairs. However, the tip-dependent variation of the splitting on each pair is in the order of 0.1 GHz, indicating that the ESR splitting was mainly contributed by the Ti-Fe interaction inherent in each pair when it formed on the surface.

**Section 2:** Eigenstates and eigenvalues of Ti-Fe pairs and Fe-Ti-Fe complex

2.1. Ti-Fe pair

To model the spin eigenstates of a Ti-Fe pair, we set up a Hamiltonian of the spin pairs

$$\widehat{H}_{\text{pair}} = D_{\text{Fe}}S_Z^2 + C_{\text{Fe}}(S_+^4 + S_-^4) + J_{\text{Ti,Fe}}\boldsymbol{S}_{\text{Ti}} \cdot \boldsymbol{S}_{\text{Fe}} - g_{\text{Ti}}\mu_B \boldsymbol{S}_{\text{Ti}} \cdot \boldsymbol{B}_{\text{ext}} - g_{\text{Fe}}\mu_B \boldsymbol{S}_{\text{Fe}} \cdot \boldsymbol{B}_{\text{ext}}, \quad \text{(S2)}$$

composed of out-of-plane MAE of Fe $D_{\text{Fe}} = -4.7$ meV (*5,19*), four-fold in-plane anisotropy of Fe $C_{\text{Fe}} = 41$ neV (*20*), Zeeman terms for Ti and Fe spins with $B_{\text{ext,x}} = 0.891$ T and $B_{\text{ext,z}} = 0.125$ T, and Ti-Fe exchange term with the coupling $J_{\text{Ti,Fe}}$ ranging 0–80 μeV, which corresponds to the peak splitting of 0–20 GHz observed from the ESR spectra (see Fig. 4). $g_{\text{Ti}}$ and $\mu_B$ are the g-factor of the Ti atom and Bohr magneton, respectively. The Hamiltonian is almost diagonal in the basis set $|m_{\text{Ti}}, m_{\text{Fe}}\rangle$, where $m_{\text{Ti}}$ and $m_{\text{Fe}}$ are magnetic quantum numbers of Ti and Fe spins, respectively (Fig. S4A). By diagonalizing the Hamiltonian, we found the four lowest eigenstates of the pair to be linear combinations of $|+1/2, +2\rangle$, $|-1/2, +2\rangle$, $|+1/2, -2\rangle$, and $|-1/2, -2\rangle$, whose eigenenergies are lower by ~ 13 meV than the other 6 eigenstates (Fig. S4B). Together with the long spin relaxation time (*20,21*), this allow us to simplify the analysis by treating the Fe as an Ising spin, and removing the MAE and four-fold in-plane anisotropy terms from the Hamiltonian. Each ESR peak originates from the resonance of the Ti spin between its two eigenstates of $m_{\text{Ti}}$ = +1/2 (+) and –1/2 (–) (*15*), and each peak in the spectra of the Ti-Fe pairs corresponds to a quasi-static Fe spin states, $|⇑\rangle$ and $|⇓\rangle$, as denoted in Fig. 1E.

2.2. Fe-Ti-Fe spin complex

To survey the eigenstates and possible ESR transitions of the Ti spin in a Fe-Ti-Fe pair shown in Fig. 5, we set up a model Hamiltonian

$$\widehat{H}_{\text{FTF}} = D_{\text{Fe}}S_{1Z}^2 + C_{\text{Fe}}(S_{1+}^4 + S_{1-}^4) + D_{\text{Fe}}S_{2Z}^2 + C_{\text{Fe}}(S_{2+}^4 + S_{2-}^4) - g_{\text{Fe}}\mu_B(\boldsymbol{S}_1 + \boldsymbol{S}_2) \cdot \boldsymbol{B}_{\text{ext}}$$
$$+ J_{\text{Ti,Fe1}}\boldsymbol{S}_{\text{Ti}} \cdot \boldsymbol{S}_1 + J_{\text{Ti,Fe2}}\boldsymbol{S}_{\text{Ti}} \cdot \boldsymbol{S}_2 - g_{\text{Ti}}\mu_B \boldsymbol{S}_{\text{Ti}} \cdot \boldsymbol{B}_{\text{ext}}, \quad \text{(S3)}$$

where two Fe spins, $\boldsymbol{S}_1$ and $\boldsymbol{S}_2$, are identical with two exchange couplings, $J_{\text{Ti,Fe1}}$ and $J_{\text{Ti,Fe2}}$, with the Ti spins, respectively. We used the same $D_{\text{Fe}}$, $C_{\text{Fe}}$, and $g_{\text{Fe}}$ as those in the Hamiltonian for a Ti-Fe pair (equation (1)) and fixed $J_{\text{Ti,Fe1}}$ to be $J_{\text{Ti,Fe}} = 1.1$ GHz as extracted from Fig. 4. We used $J_{\text{Ti,Fe2}}$ as a parameter to study a dependence of possible ESR transitions on the difference in the two Ti-Fe couplings. Similar to the case that a single Fe is coupled with the Ti spin, the Hamiltonian is almost diagonal in the basis set of the Zeeman product states of the three spins, $|m_{\text{Ti}}, m_{\text{Fe1}}, m_{\text{Fe2}}\rangle$ (Fig. S5A). The eigenstates of $\widehat{H}_{\text{FTF}}$ for the lowest 8 eigenenergies are well decoupled by the four distinct spin orientations of two Fe spins, $|⇑⇑\rangle$, $|⇓⇑\rangle$, $|⇑⇓\rangle$, and $|⇓⇓\rangle$, as shown in Fig. S5B.

**Section 3:** Model simulations of ESR transitions in Ti-Fe pairs

To survey the influence of the Ti-Fe separation on the resonance frequency of the Ti spin, we consider a simple picture that the Fe spin contributes an exchange field $\boldsymbol{B}_{\mathrm{Fe}}$ at the Ti position, as sketched in Fig. 4C. Here we use $\hat{H}_{\mathrm{pair}}$ (Eqn. S2) after removing the MAE ($D_{\mathrm{Fe}}$) and four-fold in-plane anisotropy ($C_{\mathrm{Fe}}$) terms according to the discussion in the section 2.1. The Ti-Fe interaction $J_{\mathrm{Ti,Fe}}\boldsymbol{S}_{\mathrm{Ti}} \cdot \boldsymbol{S}_{\mathrm{Fe}}$ provides an additional Zeeman energy to the Ti spin

$$\boldsymbol{B}_{\mathrm{Fe}} = (J_{\mathrm{Ti,Fe}}/\mu_{\mathrm{B}})\boldsymbol{g}_{\mathrm{Ti}}^{-1} \cdot \boldsymbol{S}_{\mathrm{Fe}}, \tag{S4}$$

where $\boldsymbol{g}_{\mathrm{Ti}}$ is the g-tensor of the Ti spin. Considering the Fe spin as a classical magnetic dipole pointing along the $\pm\hat{\boldsymbol{z}}$ direction, we take an approximation of $\boldsymbol{S}_{\mathrm{Fe}} = \langle S_{\mathrm{Fe},z} \rangle$. This simplifies the equation (S4) such that $\boldsymbol{B}_{\mathrm{Fe}} = \pm J_{\mathrm{Ti,Fe}}\langle S_{\mathrm{Fe},z}\rangle/\mu_{\mathrm{B}}g_{\mathrm{Ti},z}\hat{\boldsymbol{z}}$, as denoted in Fig. 4C by $\boldsymbol{B}_{\mathrm{Fe},\Uparrow}$ and $\boldsymbol{B}_{\mathrm{Fe},\Downarrow}$ for the Fe's $|\Uparrow\rangle$ and $|\Downarrow\rangle$ states, respectively. Diagonalization of $\hat{H}_{\mathrm{pair}}$ (Eqn. S2) yields four eigenstates, leading to two available ESR transitions, $f_{\Uparrow}$ and $f_{\Downarrow}$, of the Ti spin corresponding to two distinct Fe spin states, which determine peak splitting $\Delta f = f_{\Uparrow} - f_{\Downarrow}$. We present dependences of $f_{\Uparrow}$, $f_{\Downarrow}$, and $\Delta f$ on Ti-Fe separation in Fig. S6. Note the blue shifts of the resonance frequencies for a decreasing Ti-Fe separation, as discussed in the main text.

**Section 4:** Analysis of Rabi rates vs. tunnel conductance in Fig. 2C

<u>4.1. Two driving fields</u>

The Rabi rate ($\Omega$) of the Ti spin is directly proportional to the magnitude of the effective total driving field $\boldsymbol{B}_{1\perp}$ (see Fig. 3C), residing in the plane perpendicular to the total static field $\boldsymbol{B}_0$,

$$\Omega = \alpha B_{1\perp} \tag{S5}$$

with a proportionality constant $\alpha$. The $\boldsymbol{B}_{1\perp}$ is the vector sum of the tip-contributed ($\boldsymbol{B}_{1,\mathrm{tip}\perp}$) and Fe-contributed ($\boldsymbol{B}_{1,\mathrm{Fe}\perp}$) ones as depicted in Fig. 3C, hence its magnitude is calculated by

$$B_{1\perp} = \sqrt{B_{1,\mathrm{Fe}\perp}^2 + B_{1,\mathrm{tip}\perp}^2 + 2B_{1,\mathrm{Fe}\perp}B_{1,\mathrm{tip}\perp}\cos\phi}, \tag{S6}$$

where $\phi$ is the angle between the two vectors $\boldsymbol{B}_{1,\mathrm{tip}\perp}$ and $\boldsymbol{B}_{1,\mathrm{Fe}\perp}$. By assuming that the contribution from Fe ($\boldsymbol{B}_{1,\mathrm{Fe}\perp}$) for a given Ti-Fe separation is independent of the tunnel conductance ($\sigma_{\mathrm{tun}}$), i.e. tip-Ti distance ($d_{\mathrm{tip,Ti}}$), we show in Fig. S8B the magnitude of the total driving field ($B_{1\perp}$) as a function of that of the tip-originated driving field $B_{1\mathrm{tip}\perp}$. First, we note two special cases: (i) with $\phi = 0$ ($\boldsymbol{B}_{1,\mathrm{tip}\perp}$ and $\boldsymbol{B}_{1,\mathrm{Fe}\perp}$ are parallel.), $B_{1\perp}$, and thus $\Omega$, is maximum for the whole range of $B_{1,\mathrm{tip}\perp}$. (ii) with $\phi = 180°$ ($\boldsymbol{B}_{1,\mathrm{tip}\perp}$ and $\boldsymbol{B}_{1,\mathrm{Fe}\perp}$ are antiparallel.), $B_{1\perp}$ becomes zero at $B_{1,\mathrm{tip}\perp} = B_{1,\mathrm{Fe}\perp}$ and bounces back for $B_{1,\mathrm{tip}\perp} < B_{1,\mathrm{Fe}\perp}$. Both cases show saturation of $B_{1\perp}$ to $B_{1,\mathrm{Fe}\perp}$ at $B_{1,\mathrm{tip}\perp} = 0$, which is trivial. These two cases lead to the tunnel conductance dependence of the ESR peak amplitudes shown in Fig. 3B. For an arbitrary choice of two driving vectors, neither parallel nor antiparallel to each other, $B_{1\perp}$ shows a trend in between the two cases mentioned above and also saturates to the same point at $B_{1,\text{tip}\perp} = 0$.

### 4.2. Fit of the Rabi rates vs. tunnel conductance in Fig. 2C

Using the linear dependence of the Rabi rate of an isolated Ti spin on the tunnel conductance (gray in Fig. 2C), we can scale the conductance axis to the Rabi rate contributed only by tip ($\Omega_{\text{tip}}$), as discussed in the following. The tunnel conductance $\sigma_{\text{tun}}$ and tip-originated driving field $B_{1,\text{tip}\perp}$ can be described using the tip height ($z$) as follows (11):

$$\sigma_{\text{tun}} = \sigma_0 \, e^{-z/d_0}, \tag{S7}$$

$$B_{1,\text{tip}\perp} = c_{1,\text{tip}\perp} \, e^{-z/d_{\text{ex}}}/z, \tag{S8}$$

where $\sigma_0$, $d_0$, and $d_{\text{ex}}$ are the tunnel conductance at the point contact, decay lengths of tunneling probability, and that of exchange coupling between tip and Ti spins. Here, we take the tip height to be zero at the tip-Ti point contact. Using (S7), (S8) is transformed into

$$B_{1,\text{tip}\perp} = \frac{c_{1,\text{tip}\perp}}{d_0} \left(\frac{\sigma_{\text{tun}}}{\sigma_0}\right)^{d_0/d_{\text{ex}}} \frac{1}{\ln(\sigma_0/\sigma_{\text{tun}})}. \tag{S9}$$

Set the parts of the tunnel conductance in (S7) as $x_{\text{tip}}$ and $c_{1,\text{tip}\perp}/d_0$ as $\beta$, together with the relation in (S5), we obtain

$$\Omega_{\text{tip}}/\alpha\beta = x_{\text{tip}}, \text{ where } x_{\text{tip}} \equiv \left(\frac{\sigma_{\text{tun}}}{\sigma_0}\right)^{d_0/d_{\text{ex}}} \frac{1}{\ln(\sigma_0/\sigma_{\text{tun}})}. \tag{S10}$$

Then using (S6) and (S10), the Rabi rate ($\Omega$; y-axis of Fig. 2C) is scaled into

$$\Omega/\alpha\beta = \left(x_{\text{Fe}}^2 + x_{\text{tip}}^2 + 2\,x_{\text{Fe}}x_{\text{tip}}\cos\phi\right)^{1/2} = y, \tag{S11}$$

where $x_{\text{Fe}} \equiv B_{1,\text{Fe}\perp}/\beta$. The plots $\Omega(\sigma_{\text{tun}})$ in Fig. 2C can be transformed into $y(x_{\text{tip}})$ using (S10) and (S11), and $\alpha\beta$ can be extracted from the data for an isolated Ti using the relation in (S9).

We obtained $d_0 = 43.4$ pm by using equation (S7) with $\sigma_0 = 0.6$ mS from the point contact experiment data on a bridge-site Ti atom (22). The $d_{\text{ex}}$ was extracted using a tunnel conductance dependence of the ESR resonance frequency ($f_{\text{res}}$) measured on an isolated Ti, as shown in Fig. S9A. The linear dependence of $f_{\text{res}}$ on the tunnel conductance suggests the exchange interaction between tip and Ti as a dominant contribution to the shift of the resonance frequency. From a linear fit, we obtained $f_{\text{res}}$ of 22.485 GHz at zero conductance and calculated the shift of $f_{\text{res}}$ ($\Delta f_{\text{res}}$) at each tunnel conductance. Assuming an exponential dependence of $\Delta f_{\text{res}}$ on the tip height ($z$)

$$\Delta f_{\text{res}}(z) \propto \exp[-(z - z_0)/d_{\text{ex}}], \tag{S12}$$

we obtained the decay length of $d_{\text{ex}} = 38.2$ pm for the Ti-tip interaction. Together with the $d_0$ and $\sigma_0$ from above, we scaled the $\Omega(\sigma_{\text{tun}})$ plots in Fig. 2C into the $y(x_{\text{tip}})$ plane as shown in Fig. S10. We fit the data using (S11) with $x_{\text{Fe}}$ and $\phi$ as fitting parameters, resulting in the zero conductance Rabi rates $\Omega_0$ for the two pairs of 0.59 and 0.72 nm.

**Section 5:** Characterization of the tip used in Fig. 4

5.1. Analysis of CW- and pulsed-ESR data of an isolated Ti spin

To have a quantitative insight on the contribution of the tip spin to the measurements of the Ti-Fe pairs (Fig. 4), we performed both CW- and pulsed-ESR measurements on an isolated Ti with the same tip. In Fig. S13, we show resonance frequency ($f_{\text{ESR}}$), Rabi rate, and CW-ESR amplitude as a function of the angle ($\theta_{\text{ext}}$) of the external field. A time-independent Hamiltonian $\widehat{H}_0$ of this system can written as

$$\widehat{H}_0 = -g_{\text{Ti}}\mu_B \boldsymbol{S}_{\text{Ti}} \cdot \boldsymbol{B}_{\text{ext}} - g_{\text{Fe}}\mu_B \boldsymbol{S}_{\text{tip}} \cdot \boldsymbol{B}_{\text{ext}} + J_{\text{Ti,tip}} \boldsymbol{S}_{\text{Ti}} \cdot \boldsymbol{S}_{\text{tip}}, \tag{S13}$$

where the first two terms represent the Zeeman energies of the Ti and tip spins, and the last term does the interaction between two spins. We referred to the $g$-factor of the Ti spin as reported previously (*9*). As discussed in the main text, the tip's spin used for the measurements in Fig. 4 was well described, like the Fe spin in this work, by a classical magnetic moment of uniaxial anisotropy along a direction

$$\boldsymbol{n}_{\text{tip}} = (\sin\theta_{\text{tip}} \cos\phi_{\text{tip}}, \sin\theta_{\text{tip}} \sin\phi_{\text{tip}}, \cos\theta_{\text{tip}}), \tag{S14}$$

where $\theta_{\text{tip}}$ and $\phi_{\text{tip}}$ are the polar and azimuthal angles of $\boldsymbol{n}_{\text{tip}}$ in the Cartesian coordinates.

We choose coordinates, where the sample surface and external field are confined in the $xy$- and $xz$-plane, respectively, so that the directions of the external field and Ti spin are described by conventional definitions of the polar and azimuthal angles, $\theta_{\text{ext}}$ and $\boldsymbol{n}_{\text{tip}}(\theta_{\text{tip}}, \phi_{\text{tip}})$, as illustrated in Fig. S13A. Since the Ti-tip interaction was smaller than the Zeeman energy of the Ti by about two orders of magnitude (Fig. S14), we approximate the direction of the Ti spin to be along the external magnetic field. Hence, the external field and tip spin can be written as

$$\boldsymbol{B}_{\text{ext}} = B_{\text{ext}} \boldsymbol{n}_{\text{ext}}, \text{ where } \boldsymbol{n}_{\text{ext}} = \widehat{\boldsymbol{x}} \sin\theta_{\text{ext}} + \widehat{\boldsymbol{z}} \cos\theta_{\text{ext}}, \tag{S15}$$

$$\boldsymbol{S}_{\text{tip}} = \pm |\langle S_{\text{tip}} \rangle| \boldsymbol{n}_{\text{tip}}. \tag{S16}$$

Diagonalization of the Hamiltonian (S13) results in four eigenstates $|+\rangle|⇑\rangle$, $|-\rangle|⇑\rangle$, $|+\rangle|⇓\rangle$, $|-\rangle|⇓\rangle$, where the first and second parts of each represent Ti and tip spin states, and corresponding eigenenergies parameterized by $\theta_{\text{tip}}$, $\phi_{\text{tip}}$, and $J_{\text{Ti,tip}}|\langle S_{\text{tip}}\rangle|$. We obtain two possible ESR transitions of the Ti spin for two spin states of the tip ($|+\rangle|⇑\rangle \leftrightarrow |-\rangle|⇑\rangle$ and $|+\rangle|⇓\rangle \leftrightarrow |-\rangle|⇓\rangle$) as a function of $\theta_{\text{ext}}$. A simulation of the resonance frequency using this model with $\theta_{\text{tip}} = 65°$ and $\phi_{\text{tip}} = 75°$ is in a great agreement with the experiment (red curves in Fig. S13B). The overall $\theta_{\text{ext}}$-dependent variation of the resonance frequency ranging about 2 GHz stemmed from the anisotropy in the g-factor of the Ti spin ($g_{\text{Ti}}$) (*9*). We note a crossing of two simulation curves at about $\theta_{\text{ext}} = -25°$, across which the Zeeman energy forces the tip's spin state to flip from $|⇑\rangle$ to $|⇓\rangle$ (or vice versa).

Time-dependent perturbation theory deduces the Rabi rate ($\Omega$) of a spin $\boldsymbol{S}$ ($S = 1/2$)

$$\hbar\Omega = g\mu_B \langle + | \boldsymbol{B}_1 \cdot \boldsymbol{S} | - \rangle \tag{S17}$$

for a given time-varying magnetic field $\boldsymbol{B}_1$. Only the component of $\boldsymbol{B}_{1\text{tip}}$ perpendicular to the total static magnetic field $\boldsymbol{B}_0$ drives the ESR of the spin, leading to a dependence of the Rabi rate on the angle $\gamma$ between $\boldsymbol{B}_0$ and $\boldsymbol{B}_{1\text{tip}}$

$$\Omega(\gamma) = \Omega_{\gamma=90°} \sin\gamma = \Omega_{\gamma=90°}\sqrt{1-\cos^2\gamma}. \tag{S18}$$

Here $\gamma$ is approximately determined by the directional cosine between two unit vectors, $\boldsymbol{n}_{\text{ext}}$ and $\boldsymbol{n}_{\text{tip}}$, $\cos\gamma = \boldsymbol{n}_{\text{ext}} \cdot \boldsymbol{n}_{\text{tip}}$. The $\theta_{\text{ext}}$-dependence of the Rabi rate is then described by

$$\Omega(\theta_{\text{ext}}) = \Omega_{\gamma=90°}\sqrt{1-(\sin\theta_{\text{tip}}\cos\phi_{\text{tip}}\sin\theta_{\text{ext}} + \cos\theta_{\text{tip}}\cos\theta_{\text{ext}})^2} \tag{S19}$$

with the polar and azimuthal angles of the tip ($\theta_{\text{tip}}$, $\phi_{\text{tip}}$). Note that the tip spin flips when $\gamma$ changes across 90° or 270°, where the Rabi rate is the maximum ($\Omega_{\gamma=90°}$), leading to a continuous evolution of the Rabi rate with a periodicity of 180° in the $\theta_{\text{ext}}$ axis. With the $\theta_{\text{tip}}$ and $\phi_{\text{tip}}$ from the analysis of the resonance frequency (Fig. S13B), we simulated the $\theta_{\text{ext}}$-dependence of the Rabi rate as shown in Fig. S13C (solid orange curve), which is in good agreement with the experimental data.

The influence of the uniaxial anisotropy of the tip spin also appears in the $\theta_{\text{ext}}$-dependence of the ESR peak amplitude. We calculated a simulation curve (purple curve in Fig. S13D) using the model introduced in a previous report (*9*) with the anisotropy of our tip, $\boldsymbol{n}_{\text{tip}}(\theta_{\text{tip}}, \phi_{\text{tip}})$, extracted from the analysis of the resonance frequency (Fig. S13B). The result fits in great agreement with the experimental data, with assignments of critical angles where the ESR amplitude becomes minima when the spins of the tip and Ti are either parallel or perpendicular due to its dependence on the magnetoresistance of the tunnel junction. These angles for the perpendicular and parallel configurations of the two spins coincide with the angles for the maximum ($\Omega_{\text{max}}$) and minimum ($\Omega_{\text{min}}$) Rabi rates, a compelling evidence that the tip spin provides the driving field of the ESR on the Ti spin, which is maximum when it is perpendicular to the total static magnetic field.

5.2. CW- and pulsed-ESR data of an isolated Ti: tunnel conductance dependence

Figures S14A and B show CW- and pulsed-ESR data measured at four different tunnel conductance on the same isolated Ti atom and with the same tip used for the measurements in Fig. 4 and Fig. S13. In Fig. S14C, we show the dependence of the resonance frequency on the tunnel conductance. From a linear fit, we extrapolated the resonance frequency at zero conductance and estimated the tip-field-induced Zeeman energy of 0.008 GHz at 0.05 nS. This is smaller by ~600 times than the Ti-Fe interaction ($J_{\text{Ti,Fe}}$) of 4.91 GHz in the pair with a separation of 0.59 nm and at the field angle of 72°, as can be seen in the data in Fig. 4 and Fig. S12C where the same tunnel conductance (0.05 nS: $V_{\text{DC}}$ = 200 mV, $I_{\text{DC}}$ = 10 pA) was used. This fulfills the condition $J_{\text{Ti,Fe}} \gg J_{\text{Ti,tip}}$ in the discussion of the main text. In addition, only a noisy fluctuation was observed in the Rabi measurement at 0.1 nS (Fig. S14B), which is also a compelling evidence, supporting the negligible contribution of the tip-induced field to the data in Fig. S12C.

**Section 6:** ESR peak heights vs. tunnel conductance in Fig. 3

The contribution of the Rabi rate $\Omega$ to the ESR peak height is described by $\Omega^2 T_1 T_2/(1 + \Omega^2 T_1 T_2)$, where $T_1$ and $T_2$ are the energy relaxation time and dephasing time, respectively (*9,27*). Using a unitless Rabi rate $\Omega'_\pm \equiv \Omega_\pm \sqrt{T_1 T_2}$ as fitting parameters, our model using the Rabi rates given in Eqn. 6 excellently reproduced the tunnel conductance dependence of the ESR peak heights (solid curves in Fig. 3B). This specific behavior was observed only with certain tips, providing a collinear alignment to the Fe spin's direction, while other tips showed a similar behavior but generally yielded incomplete cancellation (Fig. S16).

**Section 7:** A model for driving ESR of a Ti spin contributed from nearby Fe

7.1. Application of 'piezoelectric driving model' to the Rabi rate contributed from Fe

The Ti-Fe interaction $J_{\text{Ti,Fe}}$ generates an inhomogeneous local magnetic field $\boldsymbol{B}_{\text{Fe}}$ at the position of the Ti atom. Then, the Ti spin in the piezo-electric motion $\Delta Z_1(t)$, induced by $\boldsymbol{E}_{\text{RF}}(t)$, can feel an effective time-varying magnetic field $\boldsymbol{B}_{1,\text{Fe}}(t)$ which is able to drive its ESR. The first order time-dependent perturbation theory yields the Rabi rate, contributed from the Fe spin,

$$\hbar \Omega_{\text{Fe}} = g\mu_B \langle +|\boldsymbol{B}_{1,\text{Fe}} \cdot \boldsymbol{S}_{\text{Ti}}|-\rangle \tag{S20}$$

for the ESR transition between two eigenstates of the Ti spin, $|+\rangle$ and $|-\rangle$. For considering the electric field in the tunnel junction, we take a parallel capacitor geometry, composed of the tip and Ag substrate as two metallic electrodes, in which the atoms have piezoelectric responses to the electric field $\boldsymbol{E}(t) = \boldsymbol{E}_{\text{DC}} + \boldsymbol{E}_{\text{RF}}(t) \approx [V_{\text{DC}} + V_{\text{RF}}(t)]/d$, (Fig. S17). Here the voltage drop in the junction occurs mostly in the vacuum region due to the larger dielectric constant of MgO than that of vacuum by the factor of ~10 (*28*), such that we take $d \approx 0.5$ nm, together with the tip-atom distance at the measurement tunnel condition when the tip's contribution was negligible (Fig. 2C; $V_{\text{DC}}$ = 100 mV, $I_{\text{DC}}$ = 10 pA) as referred to (*20,22*).

Following the discussion on so called the 'piezoelectric model' (*10*), the field induced by Fe $\boldsymbol{B}_{\text{Fe}}$ can generate a time-varying field with Eqn. S4,

$$\boldsymbol{B}_{1,\text{Fe}}(t) \approx (\partial B_{\text{Fe}}/\partial z) \cdot \Delta Z_1(t) = (\langle S_{\text{Fe},z}\rangle/\mu_B g_{\text{Ti},z}) \cdot (\partial J_{\text{Ti,Fe}}/\partial z) \cdot \Delta Z_1(t),$$

which leads to the Rabi rate contributed from the Fe spin (Eqn. S20):

$$\Omega_{\text{Fe}} = (\partial J_{\text{Ti,Fe}}/\partial z) \cdot \Delta Z_1 \cdot \sin \theta_{\text{Ti}}. \tag{S21}$$

The sine factor of the polar angle of the Ti ($\theta_{Ti}$) appears since only the component of $\boldsymbol{B}_{1,Fe}(t)$ perpendicular to the total static field ($\boldsymbol{B}_0$; i.e. the direction of the Ti spin) contributes to the Rabi rate. On assuming an isotropic Ti-Fe interaction $J_{Ti,Fe}$ (Eqn. 1), the Eqn. (S21) becomes

$$\Omega_{Fe} = -(J_{Ti,Fe}/d_{ex}) \cdot (\Delta Z_0/r) \cdot \Delta Z_1 \cdot \sin \theta_{Ti}, \qquad (S22)$$

where $d_{ex}$, $r$, and $\Delta Z_0$ are decay length of $J_{TiFe}$, Ti-Fe separation, and static displacement of the Ti atom.

Since both Ti and Fe atoms are experiencing the electric field $\boldsymbol{E}_{RF}(t)$ in this work, piezoelectric responses of both atoms should be taken account. Thus, both static ($\Delta Z_0$) and RF ($\Delta Z_1$) displacements should be measured relative to those of the Fe, such that $\Delta Z_0 = Z_{0,Ti} - Z_{0,Fe}$ and $\Delta Z_1 = Z_{1,Ti} - Z_{1,Fe}$, respectively. The static displacement $\Delta Z_0$ is the sum of contributions from the atom-surface bonding ($\Delta Z_{0,eq} = Z_{eq,Ti} - Z_{eq,Fe}$) and DC electric field ($\Delta Z_{0,DC} = Z_{DC,Ti} - Z_{DC,Fe}$). The equilibrium displacement $\Delta Z_{0,eq}$ of about 33 pm was found from our density functional theory (DFT) calculations. Resonance frequencies of atom-surface bonds for both Ti and Fe are expected to be in the range of ~ a few THz (*11*), thus we take an adiabatic approximation to the piezoelectric displacements of both atoms in our RF electric field $\boldsymbol{E}_{RF}(t)$ of a frequency range of 20–30 GHz, which allows a condition $\Delta Z_{0,DC}/V_{DC} = \Delta Z_1/V_{RF}$.

### 7.2. Density functional theory calculations for the adsorption geometry of TiH and Fe

We performed density functional theory (DFT) calculations using Quantum Espresso 7.1 (*29*) to obtain the adsorption geometry of a hydrogenated Ti (TiH) and a Fe on 2 monolayer-thick MgO on Ag(100). Projector augmented wave pseudopotentials from the PSLibrary (*30*) were used for all elements and the cutoff for the kinetic energy and charge density was set to 80 Ry and 800 Ry, respectively. Integration of the Brillouin zone was performed on a 3x3x1 regular grid of a slab-vacuum cell consisting of 4 monolayers of silver capped by 2 monolayers of magnesium oxide expanded into a 3x3 lateral cell and padded with 15 Ang of vacuum in the z-direction. Dispersive forces were treated using the Grimme-d3 method (*31*). To obtain the equilibrium geometry TiH (Fe) was added on a bridge (oxygen top) side and the whole system was relaxed until the residual forces were less than 0.01 eV/Ang. The equilibrium geometry for each system is shown in Fig. S18.

**Fig. S1.**

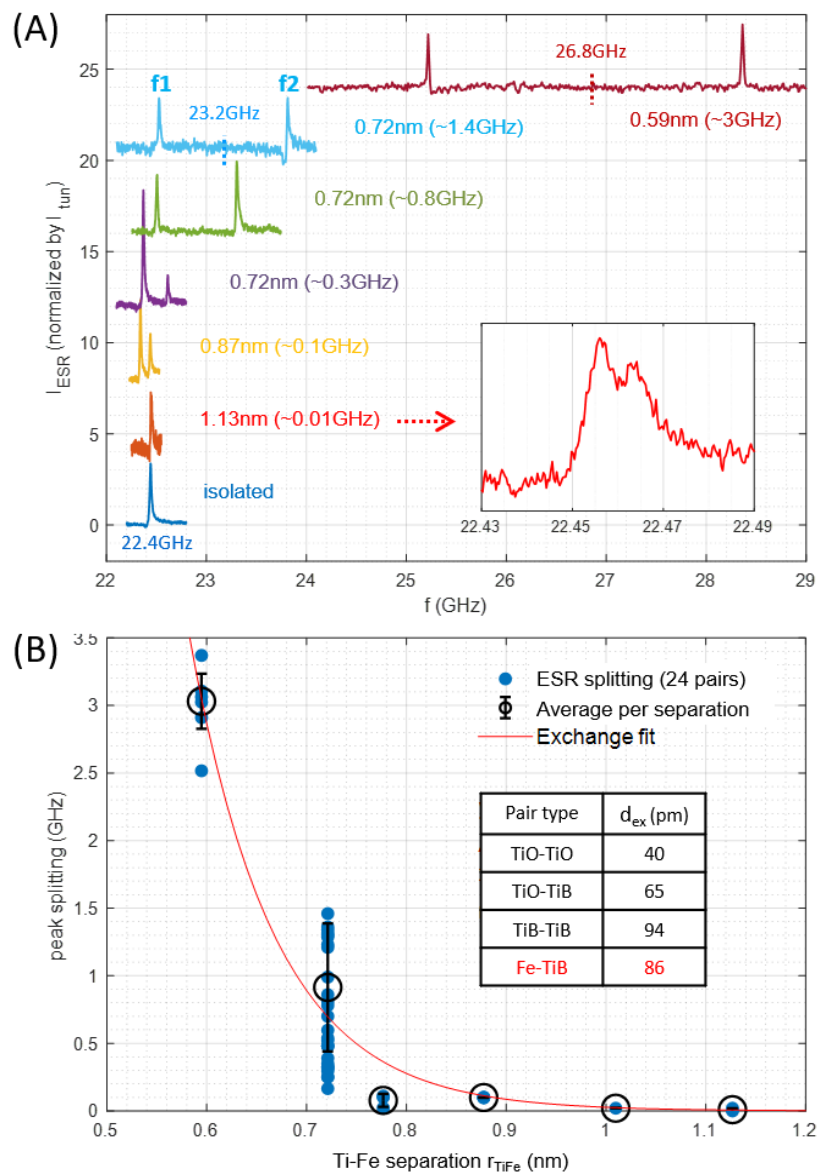

**ESR peak splitting of pairs with various Ti-Fe separations.** (**A**) Spectra measured on seven pairs of different Ti-Fe separations ($V_{RF}$ = 20 mV, $T$ = 1.2 K, $B_{ext}$ = 0.9 T, $\theta_{ext}$ = 82°). The inset is a zoom-in of the spectrum on the pair with a separation of 1.13 nm. (**B**) Dependence of peak splitting on Ti-Fe separation, extracted from 24 pairs. The red solid curve is a fit to an exponentially decaying function in the equation (S1), resulting in the decay length $d_{ex}$ of the Ti-Fe interaction shown in the table (red, inset), with those from the other spin pairs for comparison (*15,22*).

**Fig. S2.**

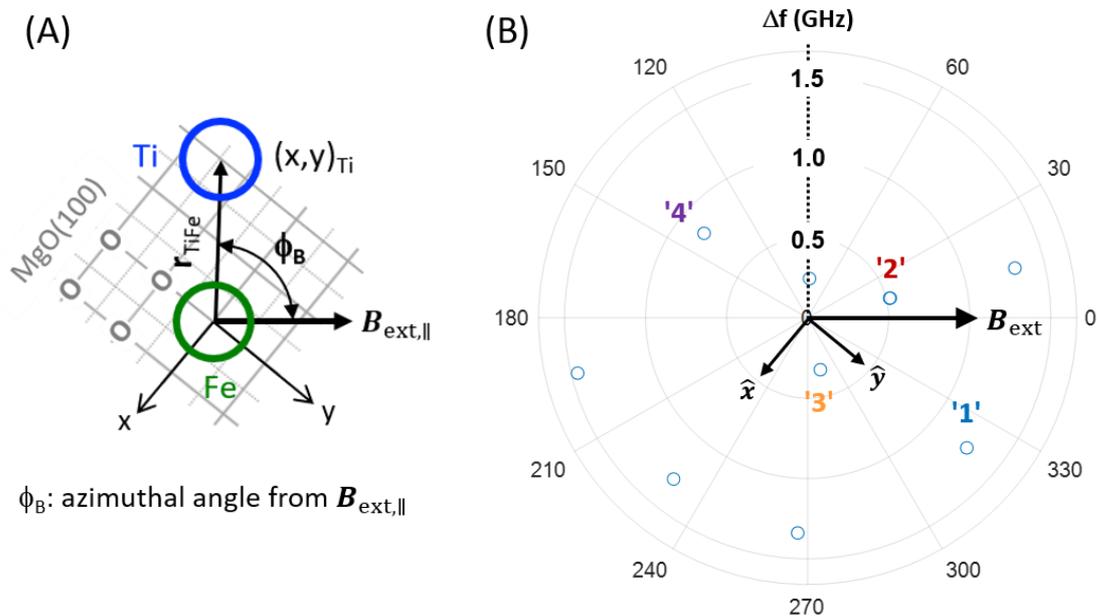

$\phi_B$: azimuthal angle from $B_{ext,\parallel}$

**Dependence of ESR peak splitting on pair orientations.** (**A**) A schematic illustrating a Ti-Fe pair with separation of 0.72 nm. The position of Ti is measured in the polar coordinate ($r_{Ti,Fe}$, $\phi_B$), measured from the position of Fe and in-plane component of the external field ($B_{ext,\parallel}$). The gray mesh represents the underlaying MgO(100) lattice. (**B**) A collection of ESR peak splitting measured on pairs with separation of 0.72 nm as a function of azimuthal angle ($\phi_B$) of Ti atom. The radius of the graph denotes the splitting ($\Delta f$) ($B_{ext}$ = 0.9 T, $\theta_{ext}$ = 82°).

**Fig. S3.**

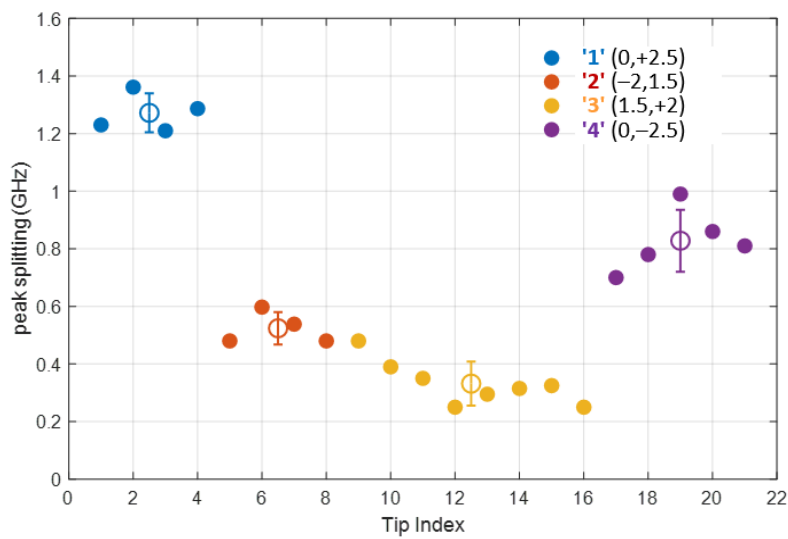

**Dependence of ESR peak splitting on different tips used.** ESR peak splitting measured from the four pairs with the same separation of 0.72 nm, denoted '1' – '4' in Fig. S2, using 21 different tips ($B_{ext}$ = 0.9 T, $\theta_{ext}$ = 82°). Note that splitting measured from pairs of 0.72 nm scatters in the range of 0.2–1.4 GHz (see also Figs. S1 and S2), however, each pair showed a much weaker dependence on the tips used, with a dispersion of ~ 0.1 GHz from its mean splitting. The open circle represents the average splitting of each pair with its error bar.

**Fig. S4.**

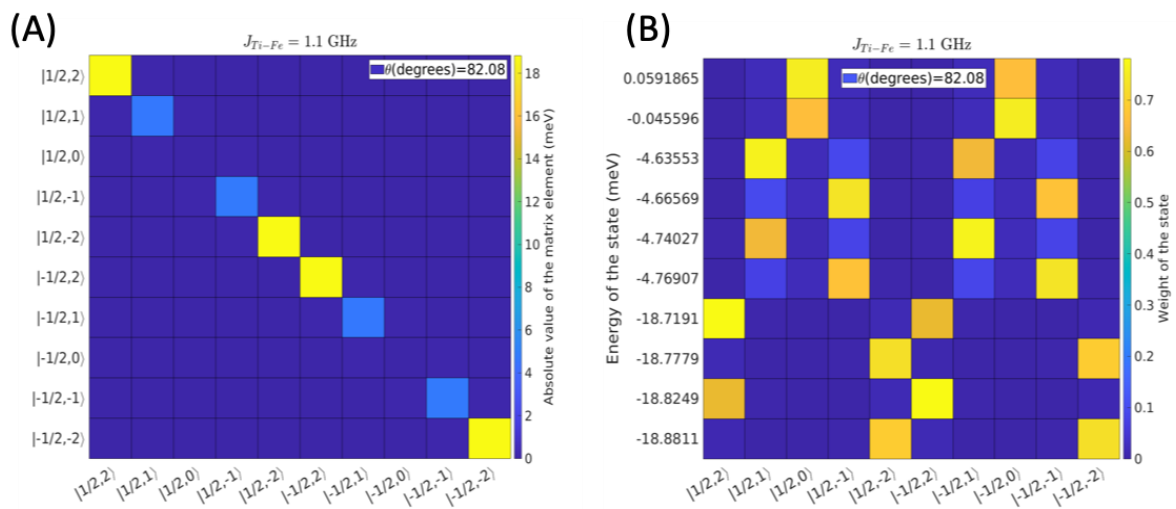

**Model Hamiltonian and eigenstates of a Ti-Fe pair.** Checkerboard representations of (**A**) the Hamiltonian Eqn. S2 for a pair of Ti-Fe separation of 0.72 nm and (**B**) its eigenstates ($B_{\text{ext}}$ = 0.9 T, $\theta_{\text{ext}}$ = 82°).

**Fig. S5.**

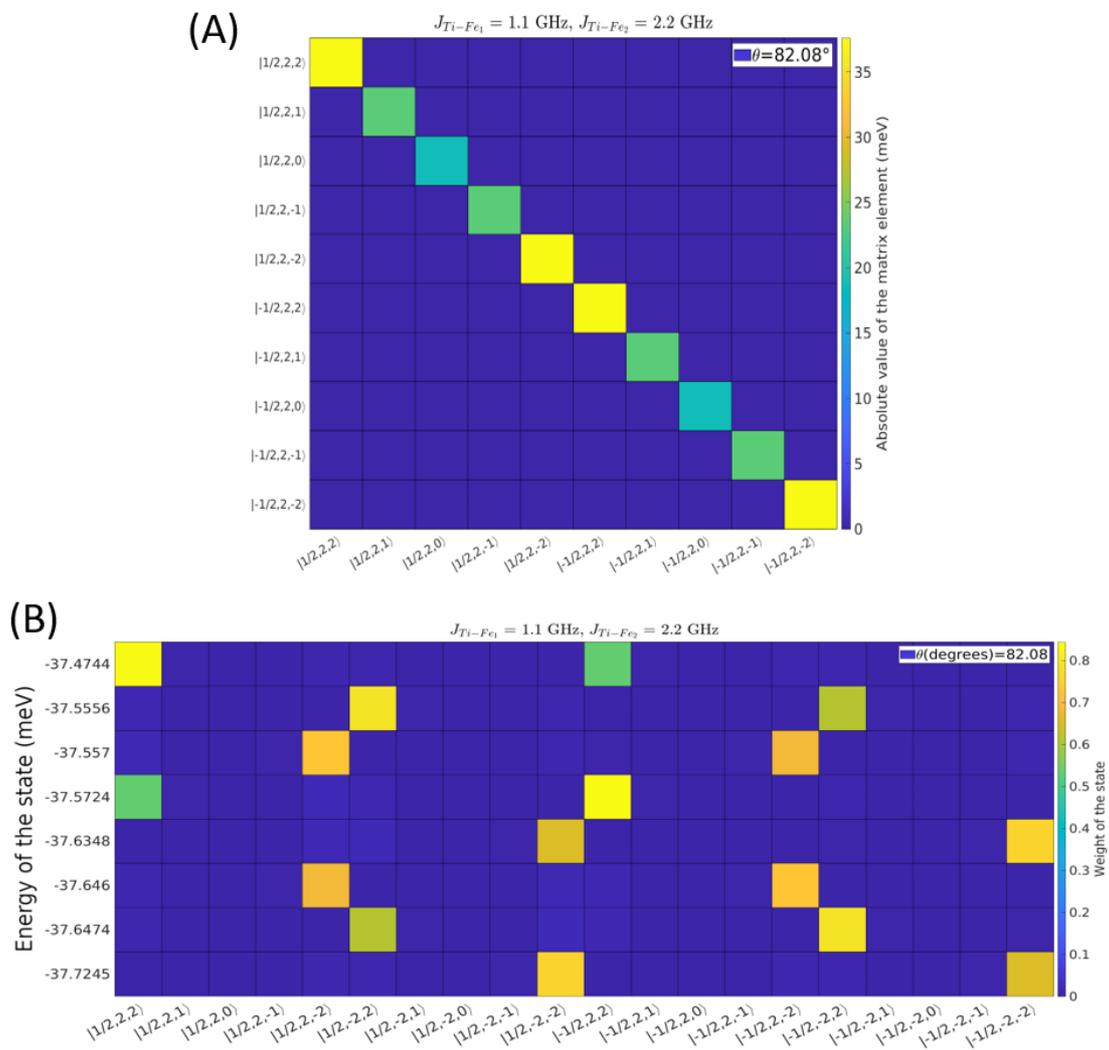

**Model Hamiltonian of a Fe-Ti-Fe spin complex.** Checkerboard representations of (**A**) subset of the Hamiltonian matrix for the first Fe (Fe₁) is up (⇑; $m_{Fe_1} = 2$) and (**B**) the eigenstates of the eight lowest energies. The spin states are labelled according to the array of three spins, $|m_{Ti}, m_{Fe1}, m_{Fe2}\rangle$ ($B_{ext}$ = 0.9 T, $\theta_{ext}$ = 82°).

**Fig. S6.**

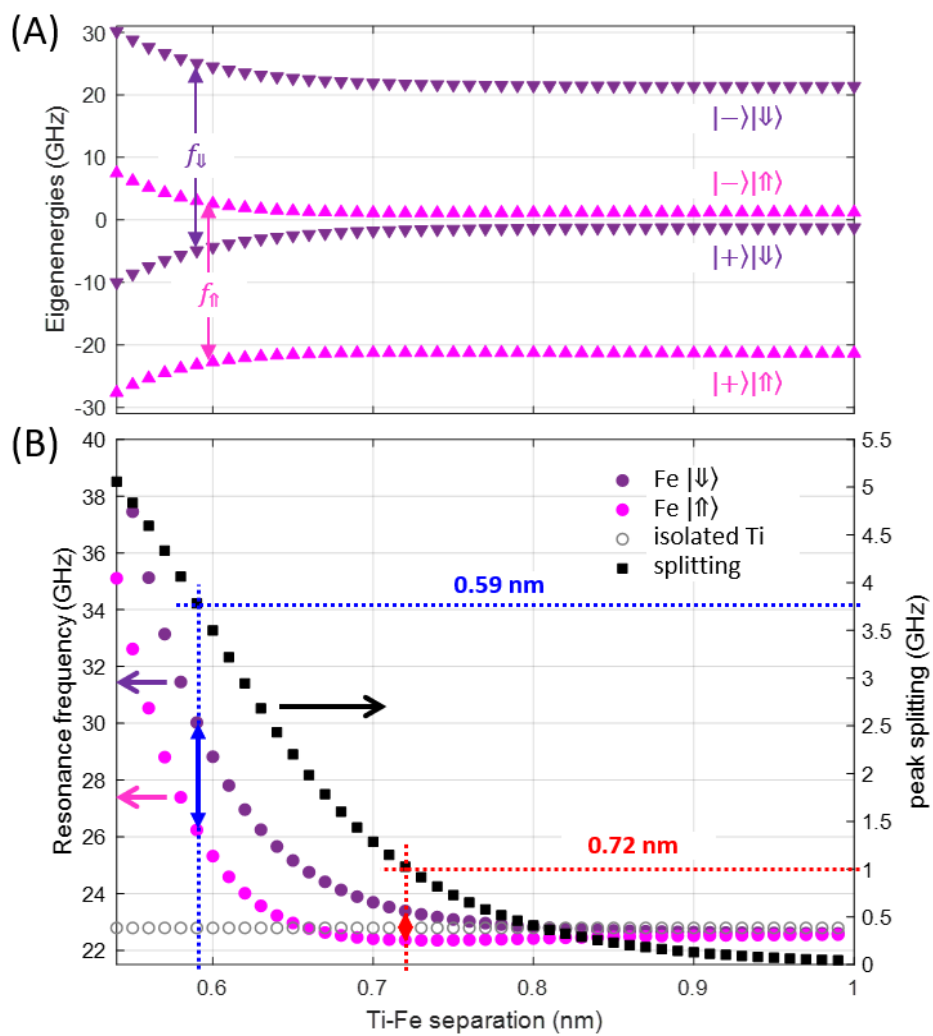

**Model simulations of ESR transitions of Ti in a Ti-Fe pair.** (**A**) Eigenstates and (**B**) peak splitting as a function of Ti-Fe separation using the model Hamiltonian $\hat{H}_{\text{pair}}$ (Eqn. S2). The dependence of $J_{\text{Ti,Fe}}$ on the Ti-Fe separation (Fig. S1B) was used. In (A), we shifted the four as-calculated eigenenergies by adding the Zeeman energies of the Fe spin for the two ESR transitions ($f_⇑$ and $f_⇓$) to be clearly visible.

**Fig. S7.**

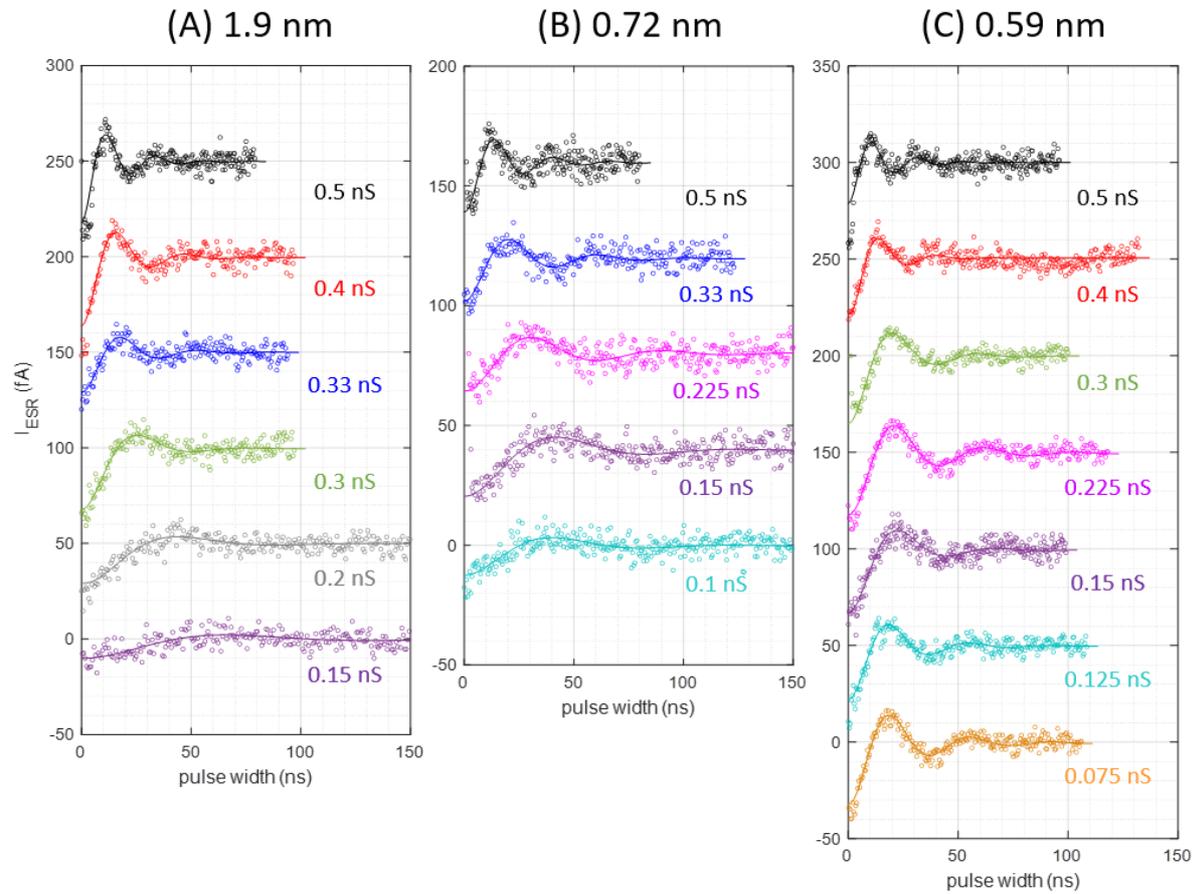

**Tunnel conductance dependence of Rabi oscillations measured on the Ti-Fe pairs shown in Figs. 1b-d.** Solid curves are the fits using an exponentially decaying sinusoidal function, resulting in the Rabi rates in Fig. 2c. ($I_{DC} = 10$ pA, $T = 1.2$ K, $B_{ext} = 0.9$ T, $\theta_{ext} = 82°$).

**Fig. S8.**

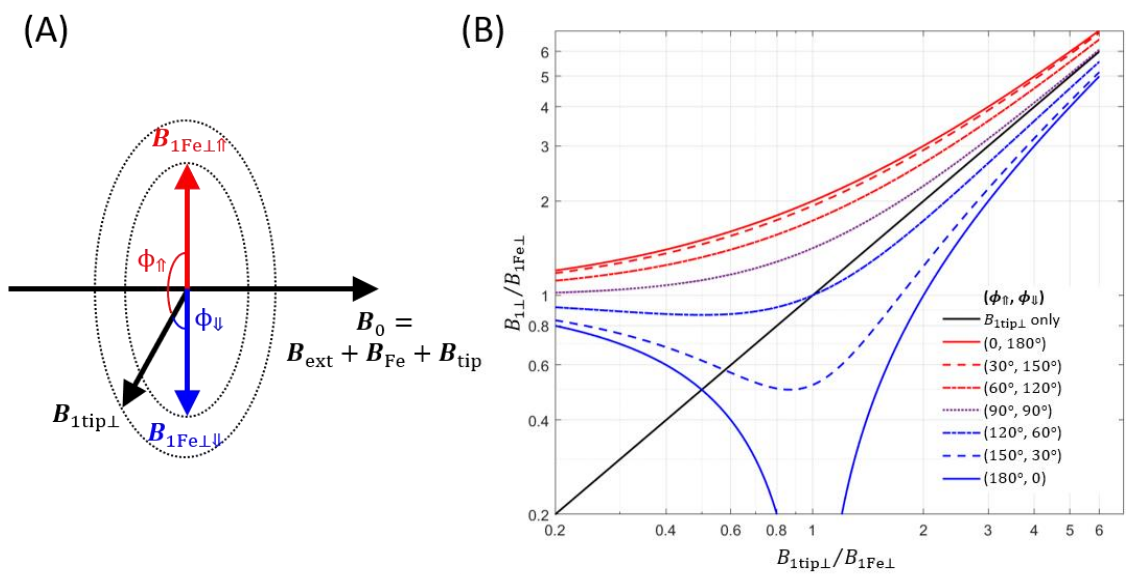

(**A**) A schematic illustration of two driving fields ($B_{1,\text{tip}\perp}$, $B_{1,\text{Fe}\perp}$) in ESR of Ti in a Ti-Fe pair. The red and blue arrows denote two Fe-induced driving fields for two quasi-static spin states of Fe, $|\Uparrow\rangle$ and $|\Downarrow\rangle$, respectively. $B_0$ is the total static field applied to the Ti spin, composed of three contributions from external, tip-induced, and Fe-induced fields. (**B**) Magnitude of total driving field ($B_{1\perp}$) following the Eqn. S6 as a function of tip-originated driving field ($B_{1,\text{tip}\perp}$), depending on the angle ($\phi_\Uparrow$ or $\phi_\Downarrow$) between $B_{1,\text{tip}\perp}$ and $B_{1,\text{Fe}\perp}$.

**Fig. S9.**

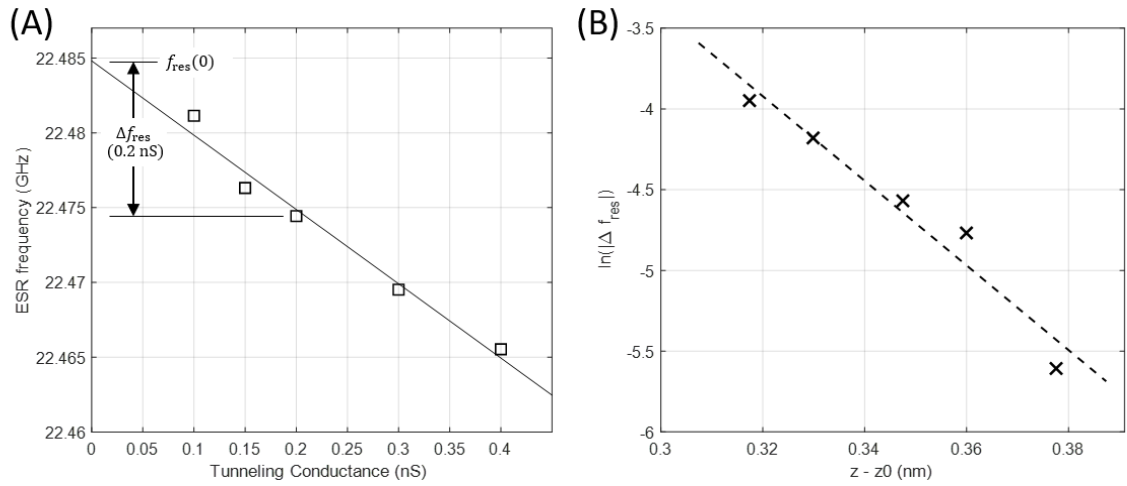

**ESR resonance frequency on an isolated Ti.** (**A**) Resonance frequency ($f_{res}$) as function of tunnel conductance. (**B**) Shift of $f_{res}$ ($\Delta f_{res}$) measured from the $f_{res}$ at zero conductance, as indicated in (A), as a function of tip height ($z$). Using as-obtained $d_0$ and $\sigma_0$ in the discussion above, we converted the measurement tunnel conductance into the tip height ($z$). The solid lines are the linear fits of the plots.

**Fig. S10.**

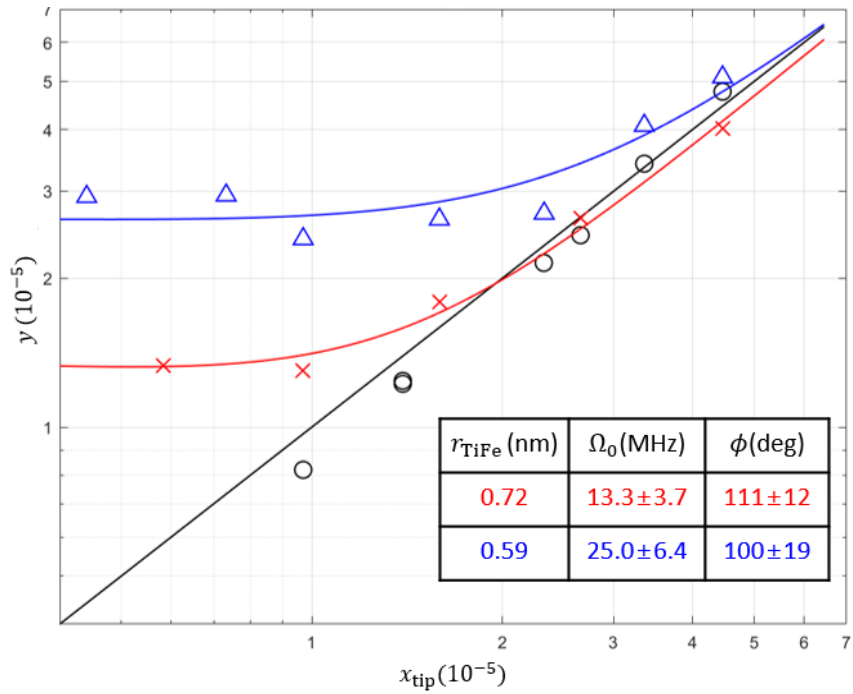

**Fits of Rabi rates vs. tunnel conductance for the three cases in Fig. 2C.** Replot the data in Fig. 2C using the model discussed in the above text with the x- and y-axis variables defined in equations (S10) and (S11). The fits resulted in the parameters $\Omega_0$ and $\phi_\Uparrow$, as shown in the inset.

**Fig. S11.**

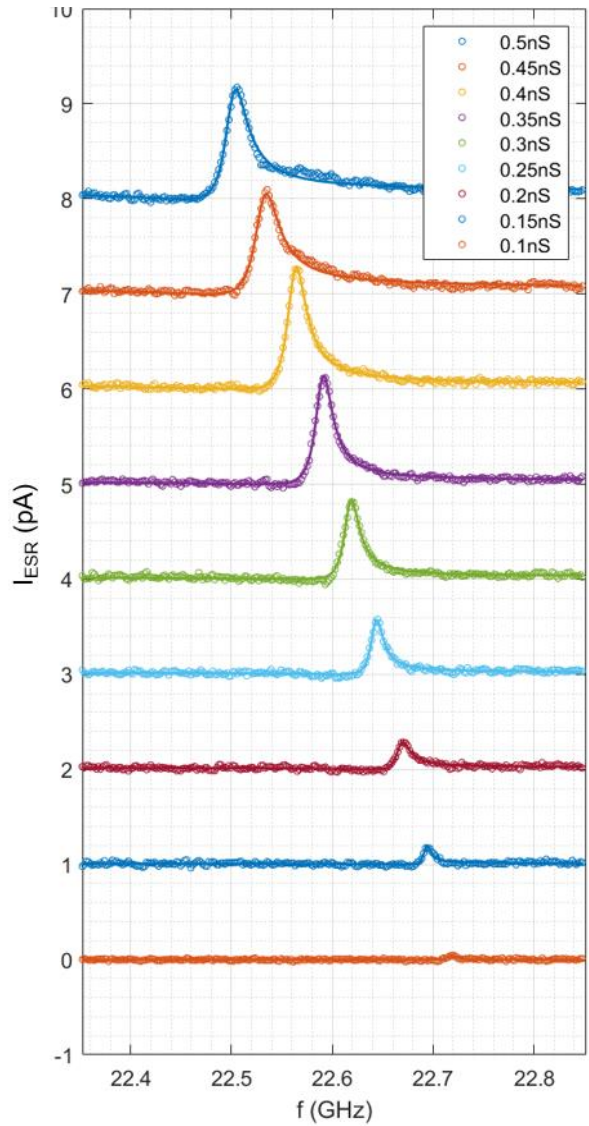

**Tunnel conductance dependence of ESR spectra of a single Ti measured with the same tip used in Fig. 3.** As the tip was retracted from the Ti spin, the peak height monotonically decreases, and it vanishes. ($V_{\text{DC}} = 50$ mV, $T = 1.2$ K, $B_{\text{ext}} = 0.9$ T, $\theta_{\text{ext}} = 82°$).

**Fig. S12.**

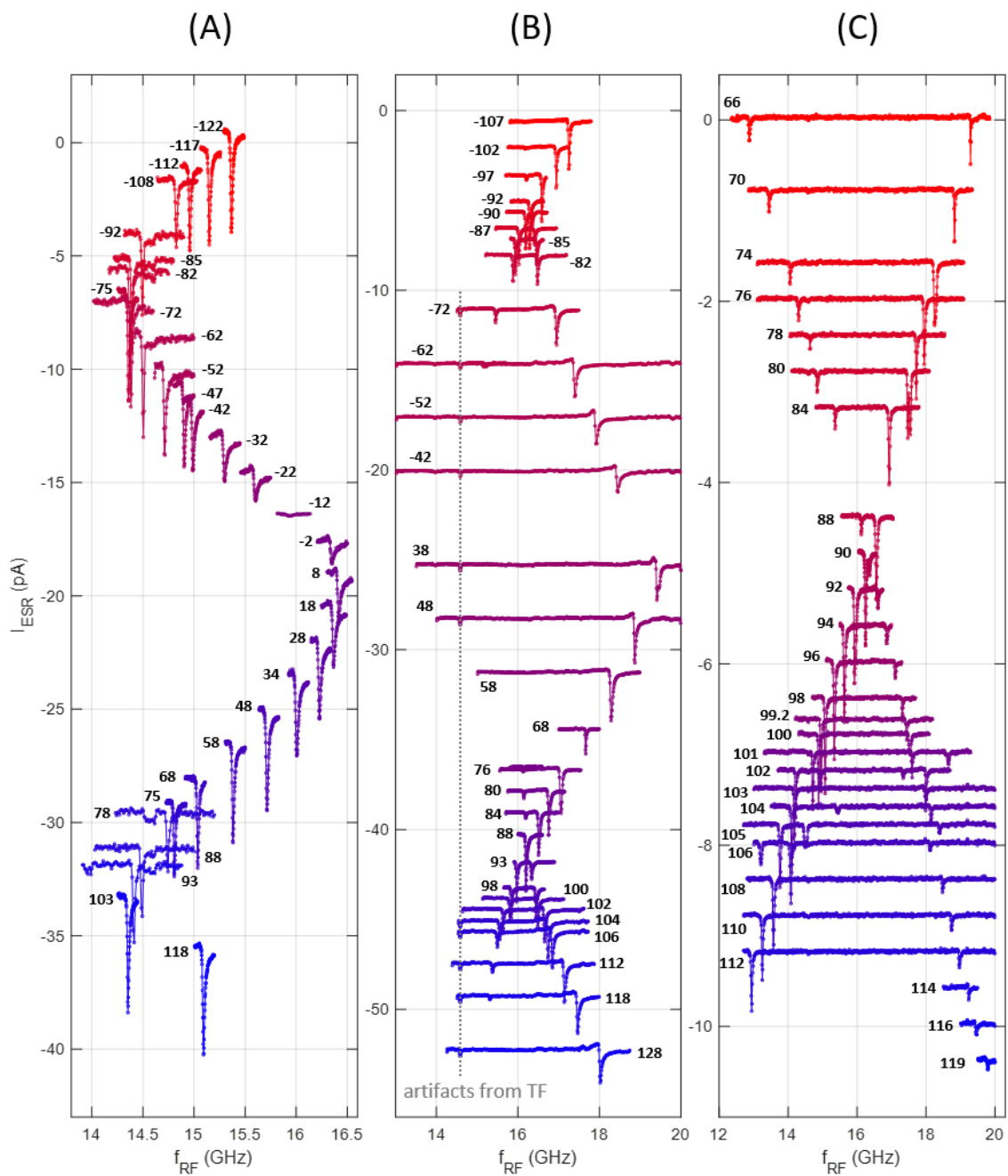

**ESR of Ti-Fe pairs with varying angle of external field ($\theta_{ext}$) for the data in Fig. 4.** ESR spectra on (**A**) an isolated Ti ($I_{tun} = 10$ pA, $V_{DC} = 30$ mV, $V_{RF} = 30$ mV) and (**B** and **C**) Ti-Fe pairs with separations of 0.72 nm ($I_{tun} = 20$ pA, $V_{DC} = 30$ mV, $V_{RF} = 30$ mV) and 0.59 nm ($I_{tun} = 10$ pA, $V_{DC} = 200$ mV, $V_{RF} = 30$ mV), respectively. Each spectrum is indicated by the polar angle, in degree (°), of the external field used in measurement and presented with a $I_{ESR}$-offset for clarity. Gray dotted line indicates artifacts from the transfer function (TF) of the RF transmission at 14.6 GHz.

**Fig. S13.**

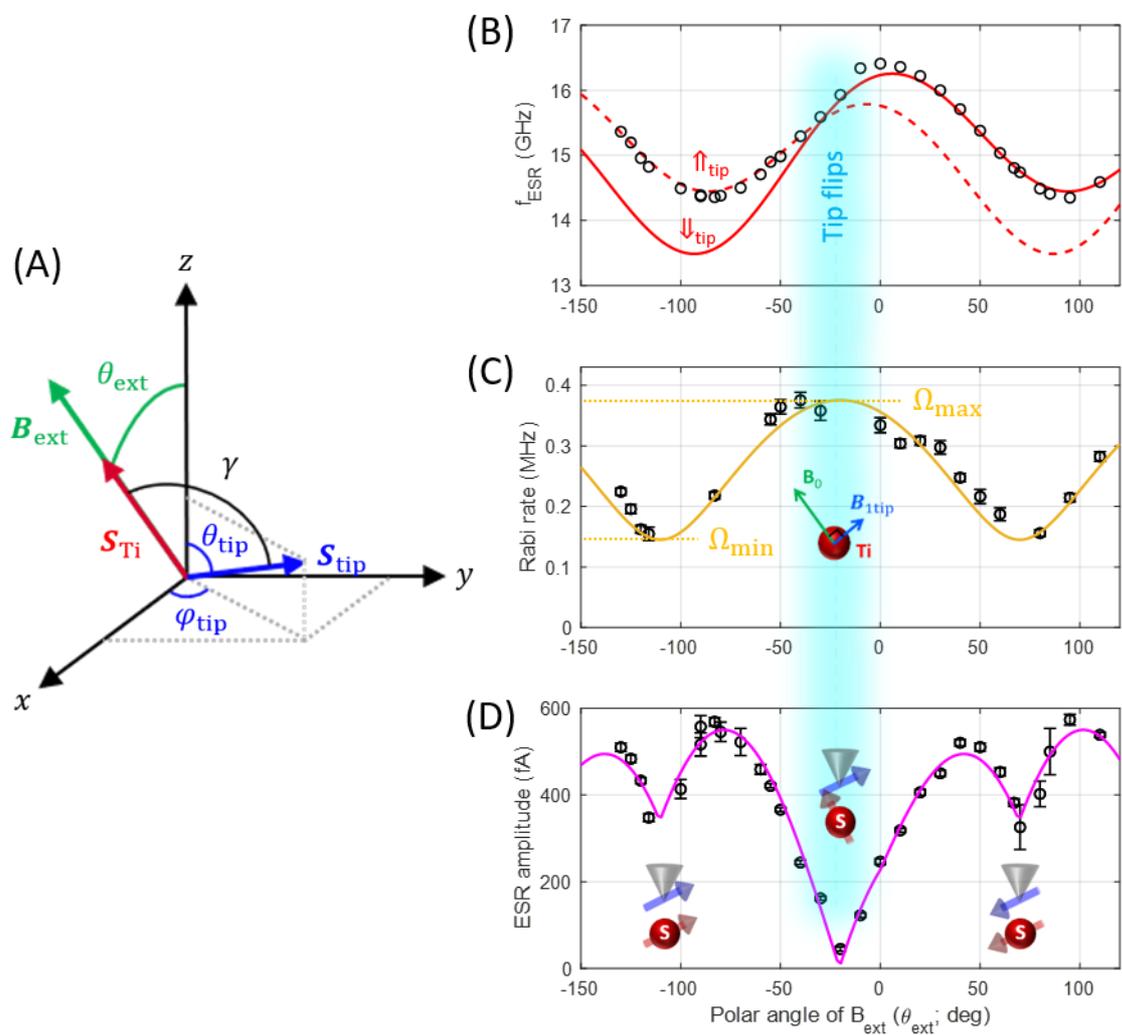

**Analysis of data from an isolated Ti spin.** (**A**) A schematic showing the definitions of the angles used in the model. Plots of (**B**) ESR frequencies, (**C**) Rabi rates, and (**D**) ESR amplitudes measured from an isolated Ti spin as a function of the polar angle of the external field ($\theta_{ext}$) ($I_{tun} = 10$ pA, $V_{DC} = 30$ mV). Solid curves are the fits using the model discussed in the text.

**Fig. S14.**

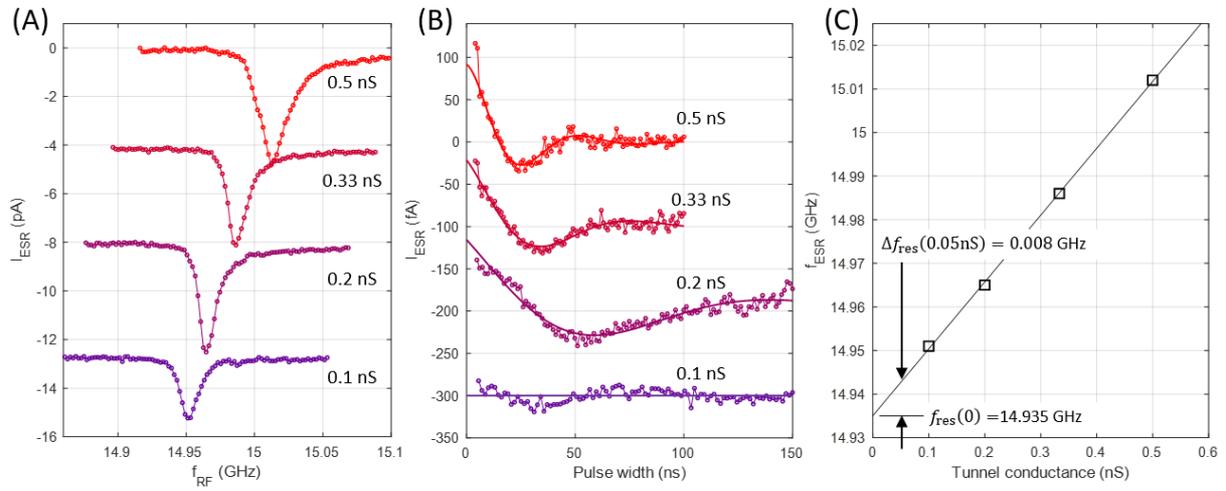

**Tunnel conductance dependence of CW- and pulsed-ESR on an isolated Ti.** (**A**) CW-ESR spectra and (**B**) Rabi oscillations measured on the same tip and Ti atom as used for the data in Fig. S13. (**C**) ESR resonance frequencies extracted from the spectra in (A). Solid line is a linear fit. ($I_{\mathrm{DC}} = 10$ pA, $T = 0.4$ K, $B_{\mathrm{ext}} = 0.6$ T, $\theta_{\mathrm{ext}} = 72°$).

**Fig. S15.**

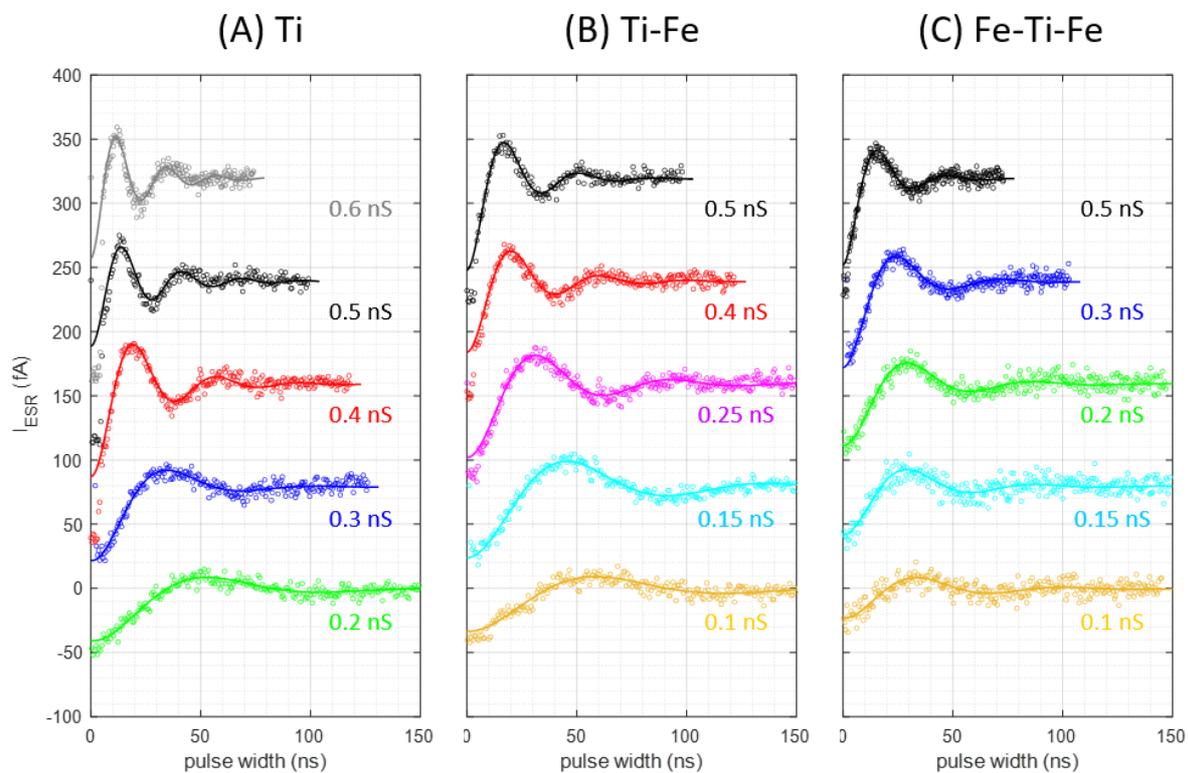

**Tunnel conductance dependence of Rabi oscillations measured on the Fe-Ti-Fe complex shown in Fig. 5.** Solid curve are the fits using an exponentially decaying sinusoidal function, giving the Rabi rates $\Omega$ shown in Fig. 5c. ($I_{DC} = 10$ pA, $T = 0.6$ K, $B_{ext} = 0.9$ T, $\theta_{ext} = 82°$).

**Fig. S16.**

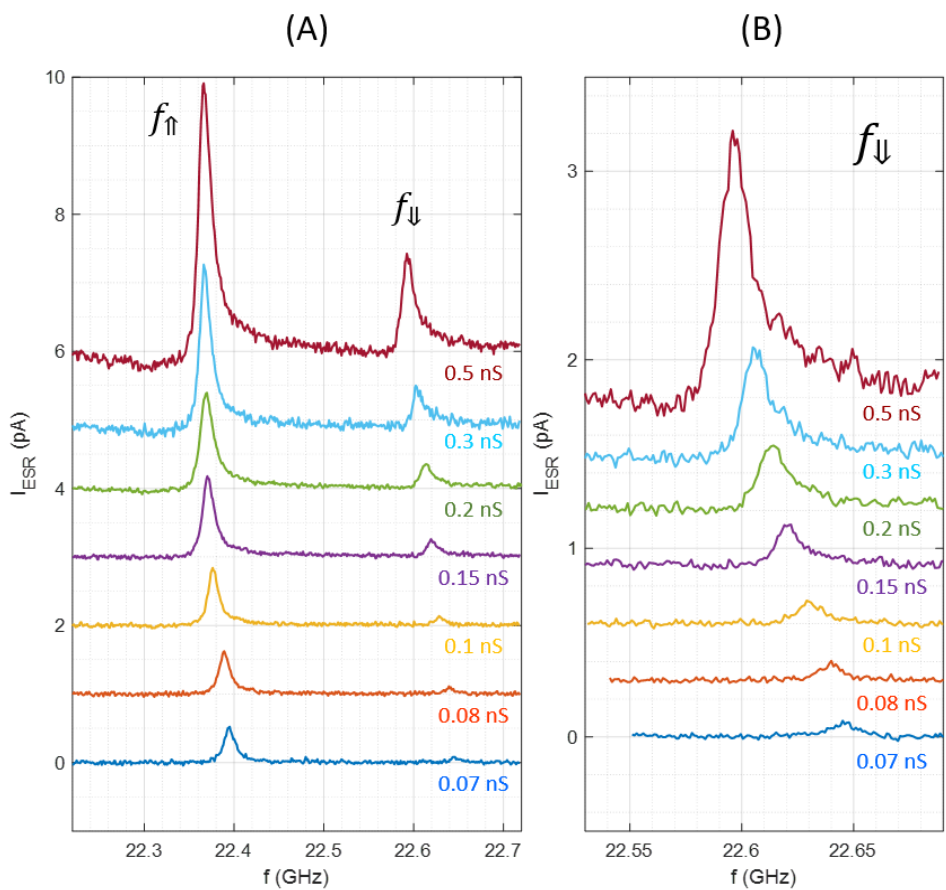

**Tunnel conductance dependence of ESR spectra on a Ti-Fe pair.** (**A**) ESR spectra showing both lower ($f_⇑$) and higher ($f_⇓$) resonances. (**B**) Zoom-in across the peak at the higher resonance ($f_⇓$). Note that both peaks showed monotonic decrease of the height as the tip was moved out from the Ti spin, in a tunnel conductance regime comparable to that in Fig. 3 ($I_{DC} = 10$ pA, $T = 1.2$ K, $B_{ext} = 0.9$ T, $\theta_{ext} = 82°$).

**Fig. S17.**

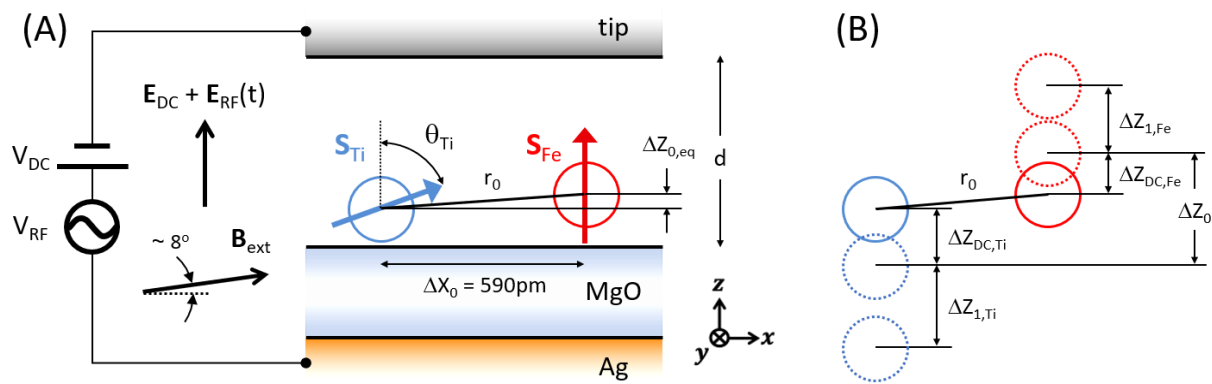

**Schematic of a Ti-Fe pair in the tunnel junction to derive piezoelectric motions of the spins.** (**A**) The geometry of the tip-substrate tunnel junction is approximated to a parallel plate capacitor, where the DC and RF biases are applied to the substrate (Ag) in series while the tip is grounded. (**B**) An illustration of piezoelectric displacements of the atoms in opposite phases (*6*) under the DC and RF electric fields shown in (A). The vertical scale for the displacements of the atoms are exaggerated compared to the horizontal scale for clarity.

**Fig. S18.**

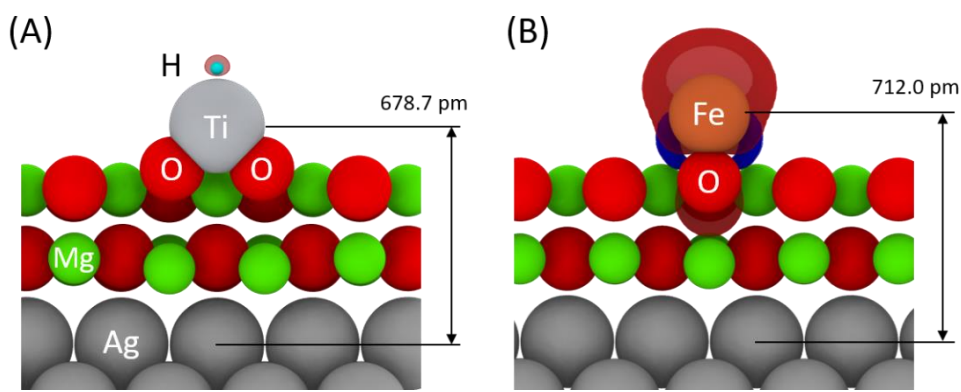

**DFT-calculated geometries of Ti (A) and Fe (B) atoms adsorbed on 2ML-thick MgO on Ag(100).**